\documentclass[aps,prd,reprint,superscriptaddress]{revtex4-2}
\usepackage{amsmath}
\usepackage{graphicx}
\usepackage{amsfonts}
\usepackage{physics}
\usepackage{amssymb}
\usepackage{bbold}
\usepackage{tikz}
\usepackage{soul}
\usetikzlibrary{arrows.meta,decorations.pathmorphing}
\usetikzlibrary{decorations.markings}
\usepackage{algorithm2e}
\usepackage[caption=false]{subfig}
\usepackage{lmodern}
\usepackage[protrusion=true,expansion=false,patch=none]{microtype}
\usepackage{etoolbox}
\AtBeginEnvironment{align}{\normalsize}
\usepackage{hyperref}
\usepackage{xcolor}

\begin{document}
\title{Traversable wormhole with double trace deformations via gravitational shear and sound channels}
\author{Fitria Khairunnisa}
\email{30223301@mahasiswa.itb.ac.id}
\affiliation{Theoretical High Energy Physics Group, Department of Physics, FMIPA, Institut Teknologi Bandung, Jl. Ganesha 10 Bandung, Indonesia.}
\affiliation{Research Center for Quantum Physics, National Research and Innovation Agency (BRIN),\\ South Tangerang 15314, Indonesia.}
\author{Hadyan Luthfan Prihadi}
\email{hady001@brin.go.id}
\affiliation{Research Center for Quantum Physics, National Research and Innovation Agency (BRIN),\\ South Tangerang 15314, Indonesia.}
    \author{M. Zhahir Djogama}
\email{30225002@mahasiswa.itb.ac.id}
\affiliation{Theoretical High Energy Physics Group, Department of Physics, FMIPA, Institut Teknologi Bandung, Jl. Ganesha 10 Bandung, Indonesia.}
\author{Donny Dwiputra}
\email{donny.dwiputra@apctp.org}
\affiliation{Asia Pacific Center for Theoretical Physics, Pohang University of Science and Technology, Pohang 37673, Gyeongsangbuk-do, South Korea.}
\affiliation{Research Center for Quantum Physics, National Research and Innovation Agency (BRIN),\\ South Tangerang 15314, Indonesia.}
\author{Freddy Permana Zen}
\email{fpzen@fi.itb.ac.id}
\affiliation{Theoretical High Energy Physics Group, Department of Physics, FMIPA, Institut Teknologi Bandung, Jl. Ganesha 10 Bandung, Indonesia.}
\affiliation{Indonesia Center for Theoretical and Mathematical Physics (ICTMP), Institut Teknologi Bandung, Jl. Ganesha 10 Bandung,
	40132, Indonesia.}
\begin{abstract}
We investigate how non-local gravitational couplings from double trace deformation between two asymptotic boundaries of an AdS$_5$ black brane can lead to the violation of the Averaged Null Energy Condition (ANEC). The first-order gravitational perturbations backreact with the background metric at second-order, creating a wormhole opening in the context of Gao-Jafferis-Wall traversable wormhole protocol. The wormhole becomes traversable in both the gravitational shear and sound channels within the hydrodynamic approximation. This shows that dynamical metric perturbations can facilitate information transfer in a purely gravitational setting, with the emergence of $G_{\text{N}}$ indicating the gravitational origin. For the shear channel, we consider three different coupling configurations, whereas for the sound channel, we vary both the speed of sound and the attenuation constant, as these parameters control the wormhole traversability. Furthermore, we obtain late-time power-law factor in the ANEC using fitting function and present a generalization that applies to both shear and sound channels. Due to its propagating nature, the sound channel exhibits late-time power-law remnants at low sound speed similar to the vector diffusive probes, but it prefers an exponential decay at higher sound speed similar to the scalar non-diffusive probes, as the power-law exponent weakened with increasing sound speed. For superluminal sound channels, the wormhole opens for an extremely brief duration at late insertion times, rendering it non-traversable.
    \end{abstract}
    \maketitle
\section{Introduction}
\indent In the AdS/CFT correspondence and holography \cite{Maldacena1999,Witten1998AdS}, the gravitational theory of anti-de Sitter (AdS) spacetime is dual to conformal field theory (CFT) defined on its boundary. As an example, a two-sided eternal AdS black hole can be described by a thermofield double state in its dual description \cite{JuanMaldacena_2003,Hartman_2013}. This  provides a powerful tool to bridge classical general relativity with quantum theory. Furthermore, this duality states the equivalence between Einstein-Rosen (ER) bridges, which arise from the general theory of relativity, and the Einstein-Podolsky-Rosen (EPR) correlation or entanglement, which originates from quantum theory. This equivalence is widely known in the literature as the ER=EPR conjecture \cite{MaldacenaSusskind,Susskind2016c,Susskind2016b}. Following this development, the study of wormholes in holography has become increasingly encouraging \cite{Gharibyan2014,Bao_2015,Remmen_2016,Dai2020,Kain_2023}.\\
\indent 
One of them is the Gao, Jafferis, and Wall (GJW) \cite{GaoJW} traversable wormhole protocol. Recently, GJW wormhole protocol has been further developed in \cite{Ahn} by considering different bulk fields. This type of traversable wormhole is realized by coupling the two asymptotic boundaries of the AdS–Schwarzschild spacetime, which induces quantum effects that give rise to negative averaged null energy. Although  the Averaged Null Energy Condition (ANEC),
\begin{equation}\label{AnecIntro}
    \int_{-\infty}^\infty T_{MN}k^M k^N d\lambda \geq0,
\end{equation}
is violated, the presence of exotic matter can be avoided in GJW protocol. The consequent ANEC violation renders the wormhole traversable (see, for example \cite{Kundu2022}, and reference therein). Here, $T_{MN}$ is stress-energy tensor, $k^M$ is a null tangent vector, and $\lambda$ represents the affine parameter along the null geodesics. They showed that quantum interactions between two nonlocally coupled CFT boundaries generate negative energy, shifting the horizon, resulting in the signal entering the horizon and reappearing in the other part of timelike spacetime. It is captured in the shift of Kruskal coordinate ($\Delta U$, as illustrated in Figure \ref{fig1}).\\
 \begin{figure*}[ht]
\centering
\begin{tikzpicture}[scale=1.1]
\begin{scope}[shift={(-4.5,0)}]
\draw[thick] (0,0) -- (0,4);
\draw[thick] (4,0) -- (4,4);

\draw[thick, decorate, decoration={snake, amplitude=0.05cm, segment length=0.1cm}]
(0,4) -- (4,4);

\draw[thick, decorate, decoration={snake, amplitude=0.05cm, segment length=0.1cm}]
(0,0) -- (4,0);

\draw[dashed] (0,0) -- (4,4);
\draw[dashed] (0,4) -- (4,0);

\node[rotate=45] at (2.6,2.3) {$U=0$};
\node[rotate=315] at (1.2,2.5) {$V=0$};

\node at (-0.5,2) {\small CFT$_L$};
\node at (4.5,2) {\small CFT$_R$};

\node at (2,-0.6) {(a) Non-traversable};
\node[align=center] at (2,4.6) {No perturbation};
\node[align=center] at (2,0.5) {$\Delta U = 0$};

\draw[
    postaction={decorate},
    decoration={
        markings,
        mark=at position 0.8 with {\arrow{>}} 
    },
    thick, blue
] (0,0.3) -- (3.7,4);

\node[rotate=45] at (0.4,1) {\small \textcolor{blue}{signal}};
\node at (-0.2,0.3) {\small $U_i$};
\node at (3.7,4.25) {\small $U_f=U_i$};
\end{scope}

\begin{scope}[shift={(2,0)}]

\draw[thick] (0,0) -- (0,4);
\draw[thick] (4,0) -- (4,4);

\draw[thick, decorate, decoration={snake, amplitude=0.05cm, segment length=0.1cm}]
(0,4) -- (4,4);

\draw[thick, decorate, decoration={snake, amplitude=0.05cm, segment length=0.1cm}]
(0,0) -- (4,0);

\draw[dashed] (0,0) -- (4,4);
\draw[dashed] (0,4) -- (4,0);

\node[rotate=45] at (0.4,1) {\small \textcolor{blue}{signal}};
\draw[
    postaction={decorate},
    decoration={
        markings,
        mark=at position 0.8 with {\arrow{>}} 
    },
    thick, blue
] (0,0.3) -- (2.85,3.15);

\draw[
    postaction={decorate},
    decoration={
        markings,
        mark=at position 0.8 with {\arrow{>}} 
    },
    thick, blue
] (3.15,2.85) -- (4,3.7);

\draw[thick, decorate, decoration={snake, amplitude=0.05cm, segment length=0.3cm}, magenta] (4,2) -- (3.15,2.85);
\draw[thick, decorate, decoration={snake, amplitude=0.05cm, segment length=0.3cm}, magenta] (2.85,3.15) -- (2,4);
\draw[thick, decorate, decoration={snake, amplitude=0.05cm, segment length=0.3cm}, magenta] (0,2) -- (2,4);

\draw[->, thick, blue] (2.85,3.15) -- (3.15,2.85);
\node at (2,1) {$\Delta U < 0$};

\node at (-0.5,0.3) {\small $U_i>0$};
\node at (4.55,3.7) {\small $U_f<0$};
\node at (-0.4,2) {\small $\mathfrak{h}_{\mu\nu}^{(0,L)}$};
\node at (4.5,2) {\small $\mathfrak{h}_{\mu\nu}^{(0,R)}$};

\node[align=center] at (2,0.5) {$\int\langle T_{VV} \rangle < 0$};

\node at (2,-0.6) {(b) Traversable};
\node[align=center] at (2,4.6) {Tensor perturbation};

\end{scope}

\end{tikzpicture}

\caption{Penrose diagrams illustrating the effect of tensor perturbations on wormhole traversability.
(a) In the absence of perturbations, null geodesics cannot cross the horizon ($\Delta U = U_f-U_i=0$), and the wormhole is non-traversable.
(b) The tensor double trace deformation backreacts on the background geometry, leading to a ANEC violation, $\int \langle T_{VV} \rangle < 0$. As a result, the null coordinate is shifted, $\Delta U < 0$, allowing signals from the left boundary to cross the horizon and reach the right boundary, thereby rendering the wormhole traversable.
}\label{fig1}
\end{figure*}
\indent The relation between the shift $\Delta U$ and the negative stress tensor can be obtained from the perturbed geodesic equation together with the linearized Einstein equations, following standard analysis (see e.g. \cite{GaoJW, Ahn}). GJW used scalar fields propagating on a three dimensional BTZ (Ba\~{n}ados, Teitelboim, Zanelli) AdS blackhole which correlate to quantum operators in the CFT boundaries. The boundary is then deformed by a non local double trace deformation \cite{Witten:2001ua,Allais:2010qq, GaoJW},
\begin{equation}
    \delta H(t)=\int d^{d-1}x \ h (t,\vec{x}) \mathcal{O}_L(-t,\vec{x})\mathcal{O}_R(t,\vec{x}).
\end{equation}
When the coupling function, $h(t,x)$, is activated, the ANEC is expected to be violated and rendering the wormhole traversable. The coupling function  connects the quantum operator $\mathcal{O}$ on the CFT right and left boundary, denoted as subscript $R$ and $L$, respectively.\\
 \indent In the GJW framework, it is important to consider ANEC instead of solely relying on the local NEC. This is because the double trace deformation is time-dependent, which results in the NEC that is not a constant. We later show in Sec. \ref{section3} that a negative NEC at some local point does not necessarily imply a negative ANEC, especially when NEC becomes positive at a later time. This should be contrasted with other works on traversable wormholes, for example \cite{nojiri2024}, where the NEC is a negative constant. In such a case, the ANEC violation is naturally satisfied. Furthermore, using the Raychaudhuri equation, one can also show that the violation of ANEC ensures that the geodesic of a null ray travelling from left to right boundaries is complete, which is a property of a traversable wormhole \cite{Kundu2022}.\\
\indent Interestingly, there is an elegant link between the traversable wormhole protocol suggested by GJW and a quantum teleportation protocol in two copies of entangled systems in a thermofield double state \cite{Maldacena_2017, Gao2021Traversable} that can be realized in a Sachdev-Ye-Kitaev (SYK) model \cite{Maldacena2016SYK}. This excitement may inspire the development of a holographic teleportation protocol that might be implemented in a tabletop experiment, offering a possible route to simulating holographic quantum gravity \cite{Schuster2022,Brown2023,Nezami2023}. Additionally, a simulation of the traversable wormhole teleportation protocol has been done in a quantum computer for a simplified sparse SYK model that is conjectured to be dual to a two dimensional holographic wormhole \cite{Jafferis2022Traversable}. Even though some of the results remain debatable \cite{Kobrin2023}, they provide valuable insight into laboratory studies of quantum aspects of gravity, especially with anticipated progress in quantum computing hardware. Hence, exploring traversable wormholes with double trace deformations induced by a broader class of coupling fields, such as the vector, tensor/gravitational, and fermionic interactions, could open new possibilities toward realizing holographic teleportation in the lab and perhaps revealing novel signatures of quantum gravitational dynamics.\\
\indent A recent work \cite{Ahn} demonstrated that a traversable wormhole can be constructed with a double trace deformation via  $U(1)$ conserved current operators. In contrast to single trace scalar operators, the $U(1)$ gauge field exhibits a power-law tail characteristic in its scrambling dynamics, reflecting a slower decay of conserved modes \cite{Cheng_2021}. It was also shown that in the hydrodynamic approximation, the nonlocal coupling between the two CFT boundaries can violate the null energy condition and render the wormhole traversable, just like the previous scalar case proposed by GJW. In that case, the retarded Green's function has a diffusive pole at $\omega\sim - i\mathcal{D}_ck^2$, where $\mathcal{D}_c$ is the diffusion constant. Furthermore, the diffusion constant appearing in the diffusive behavior of the $U(1)$ conserved current operator turns out to affect the traversability of the wormhole. They also show that the wormhole opening $\Delta U$ also has a power-law tail in the late-time limit as the reminiscent of the power-law behavior in the out-of-time order correlators (OTOC) found in \cite{Cheng_2021}.\\
\indent In this work, we construct traversable wormholes based on the GJW protocol that is included in the holographic framework theory \cite{GaoJW}. Henceforth, all instances of the term “wormhole” refer to the GJW traversable wormhole protocol. In the semiclassical approximation adopted here, Einstein's equation is written as $G_{MN}=8\pi G_\text{N}\langle T_{MN}^{\text{matter}}\rangle$, where $\langle T_{MN}^{\text{matter}}\rangle$ represents the expectation value of the stress tensor of quantum matter fields propagating on a classical background geometry \cite{Verch:1999nt}. The quantum effect of the matter fields induces negative energy gravitational shockwaves into the bulk \cite{Kundu2022}. The violation of ANEC created by the matter part ensures that the wormhole is traversable \cite{Kundu2022,GaoJW,Ahn}.\\
\indent We extend the previous works \cite{GaoJW,Ahn} by introducing the double trace deformation to a non local coupling between two tensor operators from gravitational (metric) perturbations in an AdS$_5$ black brane background (see Figure \ref{fig1}). It is shown that, in low energy, gravitational perturbations also exhibit a hydrodynamic behavior that even has richer aspects compared to the conserved current operators $U(1)$, since gravitational perturbations include scalar, vector and tensor modes (or equivalently sound, shear and scalar channels) \cite{Policastro:2002se,Herzog2003,Ge2008,MATSUO2009593} in the $h_{rN}=0$ gauge \cite{Ge2008,MATSUO2009593}. The scalar channel, $h_{xy}$, exhibits non-diffusive behavior on its retarded Green's function and is not considered in this work. The shear channels (vector modes), $h_{tx}$ and $h_{zx}$, exhibit a diffusive pole at $\omega\sim-i\mathcal{D}_Tk^2$ where $\mathcal{D}_T$ is a tensor diffusive constant that is typically half the value of the $U(1)$ conserved current operators \cite{Policastro:2002se,MATSUO2009593}. Here, we calculate $\mathcal{D}_T$ for an arbitrary black hole's blackening function $f(r)$. Furthermore, the sound channels with the remaining nonzero components have a pole at $\omega\sim\pm v_sk-i\Gamma_sk^2$, where $v_s$ is the speed of sound and $\Gamma_s$ is the attenuation constant. This represents a propagating sound wave with speed $v_s$. We explicitly calculate the null-energy condition with double trace deformation coming from both shear and sound channels and show that one could get a violation of the ANEC from both channels.\\
\indent Note that in this work, we only consider the gravitational perturbations $h_{MN}$ as the dynamical fields which obey the linearized Einstein's equation in $\mathcal{O}(h_{MN})$. The perturbations further backreact the metric background to the second order, so that the total metric components can be written as $g_{MN}=\bar{g}_{MN}+h_{MN}+\gamma_{MN}$, where $\bar{g}_{MN}$ represents the AdS black brane background and $\gamma_{MN}\sim\mathcal{O}(h_{MN}^2)$. It is this backreaction that gives rise to a wormhole opening. The Einstein's equation at the second order gives us the linearized equations of motion for $\gamma_{MN}$ sourced by the energy-momentum tensor $T_{MN}^{(2)}[h]$, which is quadratic in $h_{MN}$.\\
\indent The second-order metric perturbations $\gamma_{MN}$ deforms the null geodesic in the Kruskal coordinates so that one could relate the wormhole opening $\Delta U$ with the backreaction $\gamma_{MN}$. Due to the fact that the linearized equations of motion for $\gamma_{MN}$ is sourced by $T_{MN}^{(2)}[h]$, one can also show that the wormhole opening $\Delta U$ is proportional to $T_{VV}^{(2)}[h]$, which is nothing but the ANEC in Eq.\eqref{AnecIntro}.\\
\indent One might wonder how can we violate the ANEC from the classical solutions to the equations of motion for $h_{MN}$ to create a traversable wormhole. The ANEC associated with $h_{MN}$ can be violated once quantum effects from the double trace deformation are taken into account. In particular, one computes the quantum expectation value of $T_{VV}^{(2)}\rightarrow\langle\hat{T}_{VV}^{(2)}\rangle$ by treating the perturbation $h_{MN}\rightarrow\hat{h}_{MN}$ as a quantum operator in the interaction picture.\\
\indent We have mentioned that some of the challenges in constructing traversable wormholes are the presence of exotic matter and its stability. In this work, wormholes were constructed based on the AdS/CFT correspondence via double trace deformation. The bulk spacetime used is a black brane spacetime, which has AdS spacetime solutions in its asymptotic region. The field equation solution included a metric asymptotic expansion of the tensor field, which was considered as a perturbation and used as a source to deform the left and right CFT boundaries, resulting in nonlocal coupling between them. Then AdS/CFT correspondence is used to find a dual operator that could be obtained from the generating functional at the CFT boundary that is discussed in the context of holographic renormalization \cite{Skenderis:2002wp, deHaro:2000vlm}. This dual operator represents the expectation value of the energy-momentum tensor, which must be negative, thus constructing a traversable wormhole.\\
\indent The structure of this paper is as follows. In Section II, we write our gravitational background and the boundary condition for the tensor field in AdS$_5$/CFT$_4$. In Section III, we calculate the ANEC for both shear and sound channels. We first construct the relation between wormhole opening and the ANEC from $T_{VV}^{(2)}[h]$. Next, we find the solution to the linearized Einstein's equation in the hydrodynamic limit for the shear channel to express the tensorial diffusion constant $\mathcal{D}_T$ in terms of the background metric. We then calculate the retarded Green's and Wightman function in real space as well as the ANEC violation for both shear and sound channels. We explore both scenarios without and with the attenuation parameter. Furthermore, we investigate the traversability of the wormhole when the speed of sound becomes superluminal. In Section IV, we find a fitting model that works for both shear and sound channels simultaneously. In Section V, we briefly make comments on the the use of localized perturbations to open the wormhole. In Section VI, we summarize the results and present discussions.
\section{Gravity Setup}\label{section2}
\subsection{Black Brane Metric Background}
We begin with the general (d+1)-dimensional black brane with the metric
\begin{equation}\label{metric1}
    ds^2=g_{MN}dx^Mdx^N=g_{rr}(r)dr^2+g_{\mu\nu}(r)dx^{\mu}dx^{\nu},
\end{equation}
where $(M,N)$ indices denote the AdS $(d+1)$-coordinates and $r$ denotes the AdS radial direction. The boundary indices are given by $(\mu,\nu)=(t,x^i)$. The boundary is located at $r=0$ and the horizon is located at $r=r_H$. Near the horizon $(r/r_H=1)$, $g_{tt}$ has first order zero and $g_{rr}$ has first order pole. We also impose full rotational symmetry, $g_{ij}=g_{xx}\delta_{ij}$, in the $x^i$ direction.

In order to obtain a more well-defined view of the traversable wormhole geometry, we transform the metric to Kruskal-Szekers coordinate. Before proceeding, we define a tortoise coordinate 
\begin{equation}
    dr^*=-\sqrt{\frac{g_{rr}}{g_{tt}}}dr,
\end{equation}
so that the metric becomes
\begin{equation}
    ds^2=-g_{tt}(r^*)(dt^2-dr^{*2})+g_{xx}(r^*)\delta_{ij}dx^idx^j.
\end{equation}
Next, the Kruskal-Szekeres coordinates are defined as
\begin{align*}
    U=+e^{\frac{2\pi}{\beta}(r^*-t)},\ \ & V=-e^{\frac{2\pi}{\beta}(r^*+t)}\;\text{(left exterior)}, \\
U=-e^{\frac{2\pi}{\beta}(r^*-t)},\ \ & V=+e^{\frac{2\pi}{\beta}(r^*+t)} \;\text{(right exterior)},\\
U=+e^{\frac{2\pi}{\beta}(r^*-t)},\ \ & V=+e^{\frac{2\pi}{\beta}(r^*+t)} \;\text{(future exterior)},\\
U=-e^{\frac{2\pi}{\beta}(r^*-t)},\ \ & V=-e^{\frac{2\pi}{\beta}(r^*+t)} \;\text{(past interior)}.
\end{align*}
In these coordinates, the background metric \eqref{metric1} becomes
\begin{equation}
    ds^2=g_{UV}(UV)dU dV+g_{xx}(UV)\delta_{ij}dx^idx^j,
\end{equation}
where 
\begin{equation}
    g_{UV}=-\frac{\beta^2}{4\pi^2}\frac{g_{tt}(UV)}{UV},
\end{equation}
and $\beta=\frac{1}{T}$ is inversely related to the Hawking temperature.

To investigate the correlation function of the stress-energy tensor on the boundary, we begin by introducing small perturbations to the background metric via $g_{MN}=\bar{g}_{MN}+h_{MN}$ with $\bar{g}_{MN}$ is given by \eqref{metric1}. Since our interest lies in the boundary theory, we use the Fefferman-Graham gauge $h_{rM}=0$ for all value of $M$, eliminating any radial components of the perturbation. Thus the fluctuations now are encoded in $h_{\mu\nu}$. The perturbed metric then takes the form, 
\begin{equation}\label{metrikperturb1}
    ds^2=g_{rr}dr^2+(\bar{g}_{\mu\nu}+h_{\mu\nu})dx^\mu dx^\nu.
\end{equation}
In this case, the perturbation can be classified into three classes under spatial rotations on $xy$ plane \cite{Policastro:2002se, kovtunchannelPhysRevD.72.086009, natsuume2016adscftdualityuserguide, MATSUO2009593, Karlsson2022, Ge2008, Herzog2003}:
\begin{itemize}
    \item $h_{xy}\neq 0$, or $h_{xx}=-h_{yy}\neq0$ which include in scalar channel,
	 \item $h_{xt}$ and $h_{xz}\neq0$, or $h_{yt}$ and $h_{yz}\neq0$ which include in shear channels, and
	 \item all other elements, i.e. $h_{tt},h_{tz},h_{xx}=h_{yy},h_{zz}$ are non-zero in sound channels,
\end{itemize}
while perturbations above also refer as tensor mode, vector modes, and scalar modes respectively. The equations of motion for each classes decouple from the other ones. We consider the second and third cases since they have diffusive behavior in the (low-energy and low-temperature) hydrodynamic limit and nonzero $\omega$ poles in the Green's function.

\subsection{Boundary Conditions for the Tensor Fields}
We rewrite the bulk line element in equation \eqref{metric1} as
\begin{equation}\label{metrikbg2}
    ds^2=\frac{1}{r^2}\left(\mathfrak{g}_{rr}dr^2+\mathfrak{g}_{\mu\nu}dx^{\mu}dx^{\nu}\right),
\end{equation}
to make the near-boundary metric expansion analysis consistent with \cite{Skenderis:2002wp}. In terms of the general line element given in equation \eqref{metric1}, these metric components are written as $g_{tt}=\frac{\mathfrak{g}_{tt}}{r^2}$ and $g_{rr}=\frac{\mathfrak{g}_{rr}}{r^2}$. In this metric convention, the near-boundary limit is given by $\mathfrak{g}_{tt}(r\rightarrow0)=1$.

Now, one can treat the perturbation as a field propagating on this background with action given by,
\begin{equation}
    S=\frac{1}{2\kappa}\Bigg[\int d^{d+1}x\sqrt{g}(R[g]+2\Lambda)-\int d^dx\sqrt{\gamma}2K\Bigg],
\end{equation}
where $\kappa=8\pi G_{\text{N}}$. The second term is the Gibbons–Hawking–York boundary term that is included to ensure a well-defined variational problem under Dirichlet boundary conditions \cite{deHaro:2000vlm}.

In the framework of holographic renormalization \cite{Skenderis:2002wp,deHaro:2000vlm}, the expectation value of the boundary stress-energy tensor, $\langle T_{\mu\nu} \rangle$ is obtained from the coefficient $\mathfrak{g}_{(d)\mu\nu}$ in the near AdS boundary expansion $r\rightarrow 0$ as
\begin{equation}\label{ekspansig}
    \mathfrak{g}_{\mu\nu}(x,r)=\mathfrak{g}_{(0)\mu\nu}(x)+\dots+\mathfrak{g}_{(d)\mu\nu}r^d(x)+\dots,
\end{equation}
yielding
\begin{equation}
    \langle T_{\mu\nu} \rangle\sim\frac{\delta S_{\text{ren.}}}{\delta \mathfrak{g}_{(0)}^{\mu\nu}} \sim \mathfrak{g}_{(d)\mu\nu},
\end{equation}
while $\mathfrak{g}_{(0)\mu\nu}$ plays a role as the source to the deformation in the boundary and $S_{\text{ren.}}$ is the renormalized action \cite{deHaro:2000vlm}. In order to analyze the response of the boundary theory to small deformations of the metric, we perturb the background metric in \eqref{metrikbg2} as
\begin{equation}\label{metrikperturb2}
    ds^2=\frac{1}{r^2}[\mathfrak{g}_{rr}dr^2+(\bar{\mathfrak{g}}_{\mu\nu}+\mathfrak{h}_{\mu\nu})dx^\mu dx^\nu],
\end{equation}
where $\bar{\mathfrak{g}}$ is the unperturbed metric and $\mathfrak{h}_{\mu\nu}$ is the fluctuations with $|\mathfrak{h}_{\mu\nu}|\ll 1$. In this convention, $\mathfrak{h}_{\mu\nu}$ relates to the metric components \eqref{metrikperturb1} via $h_{\mu\nu}=\frac{1}{r^2}\mathfrak{h}_{\mu\nu}$. We then perform a similar near-boundary expansion for the perturbed metric in \eqref{metrikperturb2} yielding the boundary stress-energy tensor composed of a background and fluctuation part, $\langle T_{\mu\nu}\rangle=\langle \bar{T}_{\mu\nu}\rangle+\langle T^h_{\mu\nu}\rangle$, with ($\bar{[\dots]}$) corresponds to $\bar{\mathfrak{g}}_{\mu\nu}^{(d)}$ and $h$ corresponds to $\mathfrak{h}_{\mu\nu}^{(d)}$. 
\subsection{Double Trace Deformations}
Analogous to \eqref{ekspansig}, we now express the near-boundary expansion of the fluctuation field $\mathfrak{h}_{\mu\nu}$ as
\begin{equation}
\mathfrak{h}_{\mu\nu}=\mathfrak{h}_{\mu\nu}^{(0)}+r^2\mathfrak{h}_{\mu\nu}^{(2)}+\dots+r^d\mathfrak{h}_{\mu\nu}^{(d)}+...
\end{equation}
under Dirichlet boundary conditions, the leading term $\mathfrak{h}_{\mu\nu}^{(0)}$ acts as the source for the dual operator on the boundary. The corresponding deformation of the boundary theory action, or the generating functional, is described by
\begin{equation}
    W_h\sim \int_{\partial M} d^dx \mathfrak{h}_{\mu\nu}^{(0)}T_h^{\mu\nu},
\end{equation}
where $\mathfrak{h}_{\mu\nu}^{(0)}$ acts as the source for dual operator $T_{h,\mu\nu}$ and the expectation value of $\langle T^h_{\mu\nu}\rangle$ is proportional to $\mathfrak{h}_{\mu\nu}^{(d)}$. This has been discussed earlier in the several works \cite{Skenderis:2002wp, deHaro:2000vlm, Haro_2009}. Like the scalar and vector cases implemented by \cite{GaoJW} and \cite{Ahn}, respectively, we assume a linear relation between the source and the response is given by $\mathfrak{h}_{\mu\nu}^{(0)}=Q_{\mu\nu\alpha\beta}\mathfrak{h}^{\alpha\beta(d)}$, such that the deformation in the boundary theory is now expressed as
\begin{align}
    W_h&\sim\int_{\partial M}d^dx Q^{\mu\nu\alpha\beta}\mathfrak{h}_{\alpha\beta}^{(d)}\mathfrak{h}_{\mu\nu}^{(d)}\\\nonumber
    &\sim\int_{\partial M}d^dx Q^{\mu\nu\alpha\beta} T_{h,\alpha\beta}T_{h,\mu\nu}.
\end{align}
\indent Since there are two CFT boundaries in the maximally-extended spacetime, there are two asymptotic expansions of the gravitational fields as well. The double trace deformation introduced by \cite{GaoJW} for the traversable wormhole protocol impose a boundary condition
\begin{equation}
\mathfrak{h}_{\mu\nu}^{(0,L)}=Q_{\mu\alpha\nu\beta}\mathfrak{h}^{\alpha\beta(d,R)},\;\;\;\mathfrak{h}_{\mu\nu}^{(0,R)}=Q_{\mu\alpha\nu\beta}\mathfrak{h}^{\alpha\beta(d,L)},
\end{equation}
such that it creates a non-local interaction in the form of
\begin{equation}\label{doubletracedeformation}
    W_h\sim\int_{\partial M}d^dx Q^{\mu\alpha\nu\beta}(t,\vec{x})T^{(L)}_{h,\mu\nu}(-t,\vec{x})T_{h,\alpha\beta}^{(R)}(t,\vec{x}).
\end{equation}
The function $Q^{\mu\alpha\nu\beta}(t,\vec{x})$ represents the coupling strength between two boundaries. By setting it to be negative-valued, it creates a negative-energy shock wave that could open the wormhole, indicated by the violation of ANEC.\\
The deformation in Eq. \eqref{doubletracedeformation} is an irrelevant deformation which is different from the original GJW setup that uses relevant deformation. However, as we show in Section \ref{section3}, the irrelevant deformation which is the stress-tensor operator can render wormholes traversable in the hydrodynamic regime. We take this limit while retaining sensitivity to the leading-order effects of the deformation. The limit is discussed in the Section \ref{Section6}, related to the bound of information transfer.
\section{Wormhole Traversability}\label{section3}
In this section, we open the wormhole using the double trace deformation given by Eq. \eqref{doubletracedeformation}. We will show that, under a certain coupling, the ANEC can be violated. We relate ANEC with gravitational perturbation via Einstein Field equation. In this work, we consider gravitational perturbations which have diffusive properties such as the shear and sound channels. The Green's function has an $\omega$ pole in $\omega\sim i\mathcal{D}_Tk^2$ in the shear channel while it has a pole in $\omega\sim \pm v_sk$ in the sound channel without attenuation or $\omega\sim\pm v_sk-i\Gamma_s k^2$ with sound attenuation, where $\mathcal{D}_T$, $v_s$, and $\Gamma_s$ are the diffusive constant, the sound velocity, and sound attenuation constant respectively.\\
\indent The absolute value of the ANEC is then related to how much information can be sent through the wormhole \cite{Freivogel_2020,Ahn}. We denote this as the wormhole "traversability". A remark on the relation between the number of bits that can be transferred, $N_{\text{bits}}$, with the ANEC is later discussed in Section \ref{Section6}.
\subsection{Averaged Null Energy Condition}\label{subsectionA}
To relate ANEC violation to the wormhole opening, we calculate Einstein's equation in the quadratic expansion of the gravitational perturbation $h_{MN}$. The zeroth order of the equations of motion,
\begin{equation}
    \bar{R}_{MN}-\frac{1}{2}\bar{R} \bar{g}_{MN}+\Lambda\bar{g}_{MN}=0,
\end{equation}
give us the background solution $\bar{g}_{MN}$, which is an asymptotically-AdS black brane metric. On the other hand, the first order of the equations of motion,
\begin{align}\label{EOMEFEh}
    R^{(1)}_{MN}[h]-\frac{1}{2}(Rg_{MN})^{(1)}[h]+\Lambda h_{MN}=0,
\end{align}
give us the linearized gravitational perturbation $h_{MN}$. Therefore, the quadratic part of the equations of motion,
\begin{equation}\label{Tvvkuadrat}
    R_{MN}^{(2)}[h]-\frac{1}{2}(Rg_{MN})^{(2)}[h]\equiv-8\pi G_{\text{N}} T_{MN}^{(2)}[h],
\end{equation}
gives us the bulk energy-momentum tensor $T_{MN}^{(2)}[h]$, where $[h]$ refers that the function is $h-$dependent, which is sourced by the gravitational perturbations $h_{MN}$ and it is second ordered in $h_{MN}$, or $T_{MN}^{(2)}[h]\sim\mathcal{O}(h_{MN}^2)$.\\
\indent The energy-momentum tensor $T_{MN}^{(2)}[h]$ deforms the spacetime geometry further at the second order. Suppose that now we include the second order perturbation $\gamma_{MN}\sim\mathcal{O}(h_{MN}^2)$ to the metric such that $g_{MN}=\bar{g}_{MN}+h_{MN}+\gamma_{MN}$. The Einstein's equation at $\mathcal{O}(h_{MN}^2)$ becomes
\begin{equation}\label{EOMgamma}
    R_{MN}^{(1)}[\gamma]-\frac{1}{2}(Rg_{MN})^{(1)}[\gamma]+\Lambda\gamma_{MN}=8\pi G_{\text{N}}T_{MN}^{(2)}[h],
\end{equation}
where $R_{MN}^{(1)}[\gamma]$ is the Ricci tensor in the first order of $\gamma_{MN}$. We write the energy-momentum tensor $T_{MN}^{(2)}[h]$ as $T_{MN}^{(2)}$ hereafter for brevity. We may see the reasoning as follows. The gravitational perturbations $h_{MN}$ source further deformation to the metric at second order, denoted as $\gamma_{MN}\sim\mathcal{O}(h_{MN}^2)$ through the quadratic stress tensor $T_{MN}^{(2)}$. Then to the first order in $\gamma_{MN}$, the Einstein field equation reads \cite{GaoJW, Freivogel_2020, Ahn}
\begin{equation}
    \frac{1}{2}\Bigg(-\Lambda \gamma_{VV}+\delta^{ij}\frac{\partial_V\gamma_{ij}+V\partial_V^2\gamma_{ij}}{Vr_H^2}\Bigg)=8\pi G_{\text{N}} T_{VV}^{(2)},
\end{equation}
taking the integral over $V$, we find
\begin{equation}
    -\Lambda \int \gamma_{\text{\tiny $VV$}} dV=8\pi G_{\text{N}}\int T_{VV}^{(2)} dV.
\end{equation}
Importantly, the second-order deformation $\gamma_{MN}$ provides the necessary modification to render the wormhole traversable through its relation with the wormhole opening $\Delta U$. \\
\indent Through the geodesic equation, a null ray propagating near the horizon experiences a shift of $\Delta U$ due to the metric perturbation, which is expressed by
\begin{equation}
    \Delta U^{(1)}=-\frac{1}{g_{UV}(0)}\int dV h_{VV}.
\end{equation}
However, to the second order of metric perturbation, one need to include both metric perturbations $\{h_{MN},\gamma_{MN}\}$. Therefore, the wormhole opening at the second order is now given by
\begin{align}
    \Delta U^{(2)}=-\frac{1}{g_{UV}(0)}&\int dV \gamma_{VV}\\\nonumber&+\frac{h_{UV}}{(g_{UV}(0))^2}\int dV h_{VV},
    \end{align}
and the total wormhole opening is now given by $\Delta U = \Delta U^{(1)}+\Delta U^{(2)}$. The second term in $\Delta U^{(2)}$, as well as $\Delta U^{(1)}$, vanish in both the scalar and shear channels since $h_{VV}=0$. In the sound channel perturbation, these terms may be nonzero because $h_{tt}$ is involved; however, since $h_{tt}$ vanishes at the horizon, $h_{VV}$ can also be ignored there. Ignoring the $h_{tt}$ component, the wormhole opening at second order is directly related to the ANEC
\begin{equation}
    \Delta U^{(2)}=-\frac{1}{g_{UV}(0)}\frac{4\pi G_{\text{N}}}{3}\int dV T_{VV}^{(2)},
\end{equation}
from the second-order equation of motion \eqref{EOMgamma}. Directly calculating  $T_{VV}^{(2)}$ from $h_{MN}$ obtained by solving Eq. \eqref{EOMEFEh} classically will not make the wormhole traversable. We need to include quantum mechanical perturbations from the double trace deformation in Eq. \eqref{doubletracedeformation}.\\
\indent The linearized gravitational perturbations are then treated as a quantum field through the canonical quantization, i.e. $h_{MN}\rightarrow \hat{h}_{MN}$, and the time evolution is governed by a unitary operator $U(t,t_0)$. In this case, we can calculate the quantum expectation value of the bulk energy-momentum tensor $\langle\hat{T}_{MN}^{(2)}\rangle$ which contains the Green's function or propagator of the  $\hat{h}_{MN}$ fields since it is quadratic in $\hat{h}_{MN}$. The averaged null energy condition for traversable wormhole can then be obtained by calculating
\begin{equation}
    \text{ANEC}\sim\int_{-\infty}^\infty\langle\hat{T}_{VV}^{(2)}\rangle dV.
\end{equation}
In the following subsections, we explicitly calculate the ANEC under the double trace deformation for both shear and sound channels.
\subsection{Shear Channel}
\indent To calculate the ANEC, we use the metric in the Kruskal coordinates up to order $\sim\mathcal{O}(h_{MN})$,
\begin{align}
    ds^2=&g_{UV}dUdV+\frac{1}{r^2}\mathfrak{g}_{xx}\delta_{ij}dx^idx^j+\frac{1}{r^2}\mathfrak{h}_{\mu\nu}dx^\mu dx^\nu.
\end{align}
Furthermore, we consider the case where $\mathfrak{g}_{xx}=1$ and $g_{xx}(r)=\frac{1}{r^2}$ for simplicity. The full metric under the shear channels ($\mathfrak{h}_{tx},\mathfrak{h}_{xz}\neq0$) of gravitational perturbation is given by
\vspace{-0.5em}
\begin{align}\label{perturbedvector}
    ds^2=&-\frac{\beta^2}{4\pi^2}\frac{g_{tt}(UV)}{UV}dUdV+g_{xx}(UV)\delta_{ij}dx^idx^j\\\nonumber
    &+\frac{\beta}{2\pi UV}g_{xx}(UV)\mathfrak{h}_{tx}(U,V,z)(UdV-VdU)dx\\\nonumber
    &+2g_{xx}(UV)\mathfrak{h}_{zx}(U,V,z)dzdx.
\end{align}
\vspace{-0.5em}
\par In this setup,  $T_{VV}^{(2)}$ is defined in the low-temperature regime (large $\beta$) since we are working in low energy, hydrodynamic limit \cite{Policastro:2002se}. It is evaluated at $U=0$ and takes the form (Appendix A),
\vspace{-0.5em}
\begin{equation}\label{Tvv}
    8\pi G_{\text{N}}\langle\hat{T}_{VV}^{(2)}\rangle=-\frac{\beta^2}{32\pi^2 V^2}\langle\partial_z\mathfrak{h}_{tx}\partial_z\mathfrak{h}_{tx}\rangle+\mathcal{O}(\beta),
\end{equation}
\vspace{-0.25em}
\noindent with the $\mathcal{O}(\beta)$ terms are subleading in the hydrodynamic limit. Using the point-splitting method and ignoring the $\mathcal{O}(1)$ term, it is now written as
\begin{align}\label{ANECexpval}
    8\pi G_{\text{N}}\langle&\hat{T}_{VV}^{(2)}\rangle=- \frac{\beta^2}{32\pi^2V^2}\\\nonumber
    &\times\lim_{\textbf{x}^\prime\rightarrow \textbf{x}}\partial_z\partial_{z^\prime}\langle\mathfrak{h}_{tx}(U,V,z)\mathfrak{h}_{tx}(U^\prime,V^\prime,z^\prime)\rangle\bigg|_{U=0}.
    \end{align}
Here, $\langle\mathfrak{h}_{tx}(U,V,z)\mathfrak{h}_{tx}(U^\prime,V^\prime,z^\prime)\rangle$ is one of the Green's function of the tensor field $\mathfrak{h}_{\mu\nu}$ in the bulk, denoted as
\vspace{-0.25em}
\begin{align}
    G_{\mu\nu,\alpha\beta}(\textbf{x};\textbf{x}^\prime)=\langle\mathfrak{h}_{\mu\nu}(\textbf{x})\mathfrak{h}_{\alpha\beta}(\textbf{x}^\prime)\rangle,
\end{align}
where $\textbf{x}=(r,t,\vec{x})$ and $\textbf{x}^\prime=(r^\prime,t^\prime,\vec{x}^\prime)$.\\
\indent Due to the deformation, the bulk tensor field $\mathfrak{h}_{MN}$ evolves with the unitary time-evolution operator $U(t,t_0)=\mathcal{T}\{e^{-i\int_{t_0}^t dt_1\delta H(t_1)}\}$ in the interaction picture, where
\begin{align}
    \delta H(t_1)=\int d^3 x_1 &Q^{\mu\nu\alpha\beta}(t_1,\vec{x}_1)\\\nonumber
    &\times T_{h,\mu\nu}^{(L)}(-t_1,\vec{x}_1)T_{h,\alpha\beta}^{(R)}(t_1,\vec{x}_1),
\end{align}
is the double trace deformation. The left and right boundary is related through it. The bulk two-point function is then modified by $\delta H(t_1)$ as
\begin{align}
G_{\mu\nu,\alpha\beta}(\textbf{x};\textbf{x}^\prime)=&\langle\mathfrak{h}_{\mu\nu}^{(H)}(\textbf{x})\mathfrak{h}_{\alpha\beta}^{(H)}(\textbf{x}^\prime)\rangle\\\nonumber
    =&\langle U(t,t_0)^{-1}\mathfrak{h}_{\mu\nu}^{(I)}(\textbf{x})U(t,t_0)\\\nonumber
    &\times U(t^\prime,t_0)^{-1}\mathfrak{h}_{\alpha\beta}^{(I)}(\textbf{x}^\prime)U(t^\prime,t_0)\rangle,
\end{align}
\noindent where $(H)$ and $(I)$ in the superscript denote the Heisenberg and Interaction picture, respectively. For simplicity, the superscript $(I)$ will be omitted hereafter. We expand the time evolution operator and the bulk two-point function to the first order in $\delta H$,
\begin{align}  
U(t,t_0)&=1-i\int_{t_0}^{t}dt_1\delta H(t_1),\\
G_{\mu\nu,\alpha\beta}(\textbf{x};\textbf{x}^\prime)&=G^{(0)}_{\mu\nu,\alpha\beta}(\textbf{x};\textbf{x}^\prime)+G^{(1)}_{\mu\nu,\alpha\beta}(\textbf{x};\textbf{x}^\prime).
\end{align}
The zeroth order does not contribute to the opening of the wormhole and the violation of the ANEC. Therefore, focusing on the opening of the wormhole, we consider only the first-order Green's function as
\begin{equation}
    8\pi G_{\text{N}}\langle\hat{T}_{VV}^{(2)}\rangle=-\frac{\beta^2}{32\pi^2V^2}\lim_{\textbf{x}^\prime\rightarrow \textbf{x}}\partial_z\partial_{z^\prime}G_{tx,tx}^{(1)}(\textbf{x};\textbf{x}^\prime).
\end{equation}
\\
\indent For small $\delta H(t_1)$ perturbation, the Green's function in the first order of $\delta H(t_1)$ is given by
\begin{widetext}
\vspace{-0.5cm}
    \begin{align}\label{Gsatu}
    G_{\mu\nu,\alpha\beta}^{(1)}(\textbf{x};\textbf{x}^\prime)=&i\int_{t_0}^t dt_1\langle[\delta H(t_1),\mathfrak{h}_{\mu\nu}(\textbf{x})]\mathfrak{h}_{\alpha\beta}(\textbf{x}^\prime)\rangle+i\int_{t_0}^{t^\prime}dt_1\langle\mathfrak{h}_{\mu\nu}(\textbf{x})[\delta H(t_1),\mathfrak{h}_{\alpha\beta}(\textbf{x}^\prime)]\rangle\\\nonumber
    =&i\int_{t_0}^tdt_1d^3x_1 Q^{\lambda\rho\gamma\sigma}(t_1,\vec{x}_1)\langle[T_{h,\lambda\rho}^{(R)}(t_1,\vec{x}_1),\mathfrak{h}_{\mu\nu}(\textbf{x})]T_{h,\gamma\sigma}^{(L)}(-t_1,\vec{x}_1)\mathfrak{h}_{\alpha\beta}(\textbf{x}^\prime)\rangle+\textbf{x}\leftrightarrow \textbf{x}^\prime\\\nonumber
    =&i\int_{t_0}^tdt_1d^3x_1 Q^{\lambda\rho\gamma\sigma}(t_1,\vec{x}_1)\langle[T_{h,\lambda\rho}^{(R)}(t_1,\vec{x}_1),\mathfrak{h}_{\mu\nu}(\textbf{x})]\rangle\langle \mathfrak{h}_{\alpha\beta}(\textbf{x}^\prime)T_{h,\gamma\sigma}^{(R)}(-t_1-i\beta/2,\vec{x}_1)\rangle+\textbf{x}\leftrightarrow \textbf{x}^\prime.
 \end{align}
 \vspace{-0.5cm}
\end{widetext}
We use the large-N factorization to go from the second to the third line and we use the fact that operators in the left asymptotic boundary commutes with the ones in the right asymptotic boundary, assuming that $\mathfrak{h}_{\mu\nu}(\textbf{x})$ lives in the right region. Introducing the Wightman and retarded bulk-to-boundary propagators,
\vspace{-0.45em}
\begin{align}
K_{\mu\nu,\lambda\rho}^W(\textbf{x};t_1,\vec{x}_1)&\equiv-i\langle\mathfrak{h}_{\mu\nu}(\textbf{x})T_{h,\lambda\rho}^{(R)}(t_1,\vec{x}_1)\rangle,\\
K_{\mu\nu,\lambda\rho}^{\text{ret.}}(\textbf{x};t_1,\vec{x}_1)&\equiv-i\langle[\mathfrak{h}_{\mu\nu}(\textbf{x}),T_{h,\lambda\rho}^{(R)}(t_1,\vec{x}_1)]\rangle,
\end{align}
the first-order Green's function is then written as
\vspace{-0.5em}
\begin{align}\label{firstordergreen}
    G_{\mu\nu,\alpha\beta}^{(1)}(\textbf{x};\textbf{x}^\prime)=&
    i\int_{t_0}^t dt_1 d^3x_1Q^{\lambda\rho\gamma\sigma}(t_1,\vec{x}_1)    \\&\times K_{\mu\nu,\lambda\rho}^W(\textbf{x}^\prime;-t_1-i\beta/2,\vec{x}_1)\nonumber
    \\&\times K_{\alpha\beta,\gamma\sigma}^{\text{ret.}}(\textbf{x};t_1,\vec{x}_1)+\textbf{x}\leftrightarrow \textbf{x}^\prime.\nonumber
\end{align}
\par \noindent Here, we perform a shift $t_1\rightarrow t_1-i\beta/2$ to go from the right to the left asymptotic boundary.
\\
\indent Therefore, to calculate the ANEC in eq. \eqref{ANECexpval}, we need to calculate the Green's functions $G_{tx,tx}^{(1)}(\textbf{x};\textbf{x}^\prime)$ and $G_{zx,tx}^{(1)}(\textbf{x};\textbf{x}^\prime)$. However, we need to first specify the coupling functions $Q^{\lambda\rho\gamma\sigma}(t,\vec{x})$. There are three cases that can be considered here,
\begin{align*}
    \text{Case I: }& Q^{\lambda\rho\gamma\sigma}(t,\vec{x})=A(t)\delta_t^\lambda \delta_x^\rho \delta_t^\gamma \delta_x^\sigma,\\
    \text{Case II: }& Q^{\lambda\rho\gamma\sigma}(t,\vec{x})=B(t)\delta^\lambda_t\delta_x^\rho\delta_z^\gamma\delta_x^\sigma,\\
    \text{Case III: }&Q^{\lambda\rho\gamma\sigma}(t,\vec{x})=C(t)\delta_z^\lambda\delta_x^\rho\delta_z^\gamma\delta_x^\sigma.
\end{align*}
We also need to explicitly calculate the bulk-to-boundary propagators
\begin{align*}
    &K_{tx,tx}^{W}(\textbf{x};t_1,\vec{x}_1),\;\;\;K_{tx,tx}^{\text{ret.}}(\textbf{x};t_1,\vec{x}_1),\\
    & K_{tx,zx}^{W}(\textbf{x};t_1,\vec{x}_1),\;\;\;K_{tx,zx}^{\text{ret.}}(\textbf{x};t_1,\vec{x}_1),\\
    & K_{zx,zx}^{W}(\textbf{x};t_1,\vec{x}_1),\;\;\;K_{zx,zx}^{\text{ret.}}(\textbf{x};t_1,\vec{x}_1),
\end{align*}
and calculate these propagators by first performing a Fourier transform from a retarded bulk-to-bulk propagators in Fourier space $G_{\mu\nu,\alpha\beta}^{\text{ret.}}(r,r^\prime,\omega,k)$,
\begin{align}
  G_{\mu\nu,\alpha\beta}^{\text{ret}}(\textbf{x};\textbf{x}^\prime)=\int &\frac{d\omega}{2\pi}\frac{d^dk}{(2\pi)^d}G_{\mu\nu,\alpha\beta}^{\text{ret.}}(r,r^\prime,\omega,k)\\\nonumber
    &\times e^{-i\omega(t-t^\prime)}e^{i\vec{k}(\vec{x}-\vec{x}^\prime)},
\end{align}
then evaluating one of the bulk point to the asymptotic boundary coordinate. The Wightman propagator can be obtained from the retarded Green's function using the relation (Appendix B)
\begin{equation}\label{Wretrelation}
    G_{\mu\nu,\alpha\beta}^{W}(r,r^\prime,\omega,k)=i\frac{e^{\omega/T}}{e^{\omega/T}-1}\text{Im}G_{\mu\nu,\alpha\beta}^{\text{ret.}}(r,r^\prime,\omega,k).
\end{equation}
\subsubsection{Green's Function Calculations}\label{greenfunctionshearchannelcalc}
\indent To obtain the bulk-to-boundary propagators, we first need to solve the classical equations of motion for $\mathfrak{h}_{tx}$ and $\mathfrak{h}_{zx}$. The linearized Einstein's equation
\begin{equation}
    R_{MN}^{(1)}-\frac{1}{2}(Rg_{MN})^{(1)}-\Lambda h_{MN}=0,
\end{equation}
gives us coupled second-order differential equations. We also consider the Fourier modes of the tensor fields,
\begin{align}
    \mathfrak{h}_{tx}(r,t,z)&=\int\frac{d\omega d^3k}{(2\pi)^4}h_t(r)e^{-i\omega t}e^{i\vec{k}\cdot\vec{x}},\\
    \mathfrak{h}_{zx}(r,t,z)&=\int\frac{d\omega d^3k}{(2\pi)^4}h_z(r)e^{-i\omega t}e^{i\vec{k}\cdot\vec{x}},
\end{align}
where we assume that the tensor fields only propagate in the $z$ direction so that $\vec{k}=k\hat{z}$. The equations of motion in the Fourier modes are
\begin{align}
    \label{eomht}h^{\prime\prime}_z-h_t^\prime\partial_r&\log\bigg(\frac{g_{tt}g_{rr}}{\sqrt{|g|}g_{xx}}\bigg)\\&\nonumber-h_t\bigg(\frac{k^2g_{rr}}{g_{xx}}\bigg)
    -h_z\bigg(\frac{k\omega g_{rr}}{g_{xx}}\bigg)=0,\\
    h_{z}^{\prime\prime}-\label{eomhz}h_z^\prime\partial_r&\log\bigg(\frac{g_{rr}}{\sqrt{|g|}}\bigg)\\&\nonumber-h_z\bigg(\frac{\omega^2g_{rr}}{g_{tt}}\bigg)-h_t\bigg(\frac{k\omega g_{rr}}{g_{tt}}\bigg)=0,\\&
\label{eomhthz}\;\;\;\;\;\;\;\;\;\;\;\;\;\;\;\;\;\;\;\;\;\; \omega g_{xx}h_t^\prime-kg_{tt}h_z^\prime=0.
\end{align}
From equation \eqref{eomht}, we can solve for $h_z(r)$ and obtain
\begin{equation}
    h_z=\frac{g_{xx}}{k\omega g_{rr}}\bigg\{h_t^{\prime\prime}-h_t^\prime\partial_r\log\bigg(\frac{g_{tt}g_{rr}}{\sqrt{|g|}g_{xx}}\bigg)-h_t\bigg(\frac{k^2g_{rr}}{g_{xx}}\bigg)\bigg\}.
\end{equation}
Inserting this to \eqref{eomhthz}, we get equation of motion for the perturbation,
\begin{align}
    h_t^{\prime\prime\prime}&-h_t^{\prime\prime}\partial_r\log\bigg(\frac{\sqrt{|g|}g_{rr}}{g_{xx}^5}\bigg)-g_{rr}\bigg[\omega^2g^{tt}+k^2g^{xx}\\\nonumber&+g^{xx}\partial_r\bigg(\frac{g_{xx}^5}{\sqrt{|g|}g_{rr}}\partial_r\bigg(\frac{\sqrt{|g|}}{g_{xx}^4}\bigg)\bigg)\bigg]h_t^\prime=0.
\end{align}
\indent We now proceed to find the solution for the equation of motion. For simplicity, we consider a $(4+1)$-dimensional AdS black hole spacetime in the $(t,u,\vec{x})$ coordinates,
\begin{equation}
    ds^2=\frac{1}{u}\big(-f(u)dt^2+dx^2+dy^2+dz^2\big)+\frac{du^2}{4u^2f(u)},
\end{equation}
where $f(u)$ satisfy $f(u_H)=0$, with $u_H$ be the horizon's radius and $f(0)=1$ from the AdS boundary condition. This metric can be obtained by defining $u=r^2$ in Eq. \eqref{metrikbg2}. In this case, we have
\begin{equation}
    g_{rr}(r)\rightarrow g_{uu}(u)=\frac{1}{4u^2f(u)},\;\;\;g_{xx}=\frac{1}{u},\;\;\;\sqrt{|g|}=\frac{1}{2u^3}.
\end{equation}
The equation of motion for $h_t(u)$ then becomes
\begin{align}
    h_t^{\prime\prime\prime}(u)&-h_t^{\prime\prime}(u)\partial_u\log(u^2g_{uu})\\\nonumber-&ug_{uu}\bigg(\frac{\omega^2g^{tt}}{u}+k^2\bigg)h_t^\prime+\frac{1}{u}\partial_u\log(u^2g_{uu})h_t^\prime=0.
\end{align}
We also assume that the black hole is non-extremal so that $g_{uu}(u)$ has an order-1 pole near the horizon,
\begin{equation}
    g_{uu}(u)\approx\frac{1}{4u_H^2f^\prime(u_H)(u-u_H)}.
\end{equation}
\indent The solutions with infalling boundary condition at the horizon in the hydrodynamic limit with $(\omega,k^2)\ll1$ are given by (Appendix C)
\begin{align}\label{solhzprime}
h_t^\prime(u)&=\frac{k^2h_t^0+\omega kh_z^0}{H_1(0)(i\omega-\mathcal{D}_Tk^2)}\left(1-\frac{u}{u_H}\right)^{-i\omega/\nu}\\\nonumber
    &\;\;\;\;\;\;\;\times\bigg(u-i\omega H_1(u)+k^2H_2(u)+\dots\bigg),\label{solhtprime}\\
    h_z^\prime(u)&=-\frac{\omega kh_t^0+\omega^2h_z^0}{f(u)H_1(0)(i\omega-\mathcal{D}_Tk^2)}\left(1-\frac{u}{u_H}\right)^{-i\omega/\nu}\\\nonumber
    &\;\;\;\;\;\;\;\times\bigg(u-i\omega H_1(u)+k^2H_2(u)+\dots\bigg),
\end{align}
where $\dots$ are the terms with $\mathcal{O}(\omega^2,\omega k^2,k^4)$ and higher order. Here, $H_1(u)$ and $H_2(u)$ are defined as
\begin{align}\label{h2h1}
    H_1(u)=&\frac{u}{\nu}\int_{u}^{u_H}\bigg(\frac{u_H^2|F^\prime(u_H)|}{u_1^2f(u_1)}-\frac{1}{u_H-u_1}\bigg)du_1,\\[10pt]\nonumber 
    H_2(u)=&\frac{u}{8}\int_{u_H}^u\frac{(u_1^2-u_H^2)}{u_1^2f(u_1)}du_1,
\end{align}
and
\begin{equation}
    \mathcal{D}_T\equiv\frac{H_2(0)}{H_1(0)},
\end{equation}
is the diffusion constant for the tensor perturbations. For a Scwarzschild-AdS$_5$ black hole with $f(u)=1-(\frac{u}{u_H})^2$, we recover $\mathcal{D}_T=\frac{\beta}{4\pi}$, which is consistent with \cite{Policastro:2002se}.\\
\indent We can switch from $(t,u,\vec{x})$ coordinates to the tortoise coordinates $(t,r^*,\vec{x})$ using the relation
\begin{equation}
    r^*(u)=\int^{u}\sqrt{-\frac{g_{uu}(u_1)}{g_{tt}(u_1)}}du_1.
\end{equation}
Near the horizon, following relation applies
\begin{equation}
    r^*(u)\approx\frac{1}{\nu}\log|u_H-u|.
\end{equation}
Thus, we have the near horizon solutions in the form of
\begin{align}\label{solutionsapprox}
    h^\prime_t(r^*)\approx& \frac{k^2h_t^0 e^{-i\omega r^*}u_H^{i\omega/\nu+1}}{H_1(0)(i\omega-\mathcal{D}_Tk^2)},\\\nonumber
    h^\prime_z(r^*)\approx&- \frac{\omega^2h_z^0 e^{-i\omega r^*}u_H^{i\omega/\nu+1}}{f(r^*)H_1(0)(i\omega-\mathcal{D}_Tk^2)}.
\end{align}
\indent Furthermore, we can calculate the bulk-to-bulk retarded Green's function in Fourier space
\begin{align}
    G_{\mu\nu,\alpha\beta}^{\text{ret.}}(r^*,r^{*\prime},\omega,k)&=-i\int dt d^3 xe^{i\omega t-i\vec{k}\cdot\vec{x}}\Theta(t)\\\nonumber
    &\times\langle[T_{\mu\nu}(r^*,t,x),T_{\alpha\beta}(r^{*\prime},0,0)]\rangle,
\end{align}
analogous to the calculations performed with the $U(1)$ Maxwell gauge field \cite{Ahn}. One can do this by first solving the homogeneous equations of motion for the respective fields in the hydrodynamic limit which has a diffusive pole. In our case, the solutions are given by Eq. \eqref{solutionsapprox}. Then, the bulk-to-bulk retarded Green functions ansatz in Fourier space that are also consistent with \cite{Policastro:2002se,Ge2008,Cheng_2021} are given by
\begin{align}
    G_{tx,tx}^{\text{ret.}}(r^*,r^{*\prime},\omega,k)&\approx\frac{\mathcal{D}_T^2k^2e^{-i\omega(r^*-r^{*\prime})}}{(i\omega-\mathcal{D}_Tk^2)},\\
    G_{tx,zx}^{\text{ret.}}(r^*,r^{*\prime},\omega,k)&\approx i\frac{\mathcal{D}_T^2\omega ke^{-i\omega(r^*-r^{*\prime})}}{(i\omega-\mathcal{D}_Tk^2)},\\
    G_{zx,zx}^{\text{ret.}}(r^*,r^{*\prime},\omega,k)&\approx\frac{\mathcal{D}_T^2\omega^2e^{-i\omega(r^*-r^{*\prime})}}{(i\omega-\mathcal{D}_Tk^2)},
\end{align}
up to some proportionality constant. The factor $\mathcal{D}_T^2$ is added to ensure that the Green's functions are dimensionally correct. The difference between our calculations and the previous works is that we generalize the diffusion constant $\mathcal{D}_T$ for arbitrary function $f(u)$. \\ 
\indent Now, we want to calculate the bulk-to-boundary correlators in real space by inverse Fourier transform
\begin{align}
    G_{tx,tx}(\textbf{x};\textbf{x}^\prime)=\int\frac{d\omega}{2\pi}&\frac{d^3k}{(2\pi)^3}G_{tx,tx}(r^*,r^{*\prime},\omega,k)\\\nonumber
    &\times e^{-i\omega(t-t^\prime)}e^{i\vec{k}\cdot(\vec{x}-\vec{x}^\prime)}.
\end{align}
For $G_{tx,tx}^{\text{ret.}}$, it is given by
\begin{align}
    G_{tx,tx}^{\text{ret.}}(\textbf{x};\textbf{x}^\prime)=&\int \frac{d\omega}{2\pi}\frac{d^3k}{(2\pi)^3}\mathcal{D}_T^2k^2\frac{e^{-i\omega(r^*-r^{*\prime})}}{i\omega-\mathcal{D}_Tk^2}\\\nonumber
    &\times e^{-i\omega(t-t^\prime)}e^{i\vec{k}\cdot(\vec{x}-\vec{x}^\prime)}\\\nonumber
    =&\int\frac{d^3k}{(2\pi)^3}\mathcal{D}_T^2k^2e^{i\vec{k}\cdot(\vec{x}-\vec{x}^\prime)}\\\nonumber
    &\times\bigg(\int_{-\infty}^\infty\frac{d\omega}{2\pi}\frac{(V/V^\prime)^{-i\omega/2\pi T}}{i\omega-\mathcal{D}_Tk^2}\bigg)\\\nonumber
    =&-\int\frac{d^3k}{(2\pi)^3}\frac{\mathcal{D}_T^2k^2e^{i\vec{k}\cdot(\vec{x}-\vec{x}^\prime)}}{(V/V^\prime)^{\frac{\mathcal{D}_Tk^2}{2\pi T}}}\Theta(V-V^\prime),
\end{align}
where $\Theta(x)$ is the Heaviside step function obtained  from closing the $\omega$ contour down and pick up a pole in the lower-half plane. Using the relation in Eq. \eqref{Wretrelation}, the Wightman Green's function is given by
\begin{align}
    G_{tx,tx}^W(\textbf{x};\textbf{x}^\prime)=&-\frac{i}{4}\Theta(V-V^\prime)\\\nonumber
    &\times\int\frac{d^3k}{(2\pi)^3}\frac{\mathcal{D}_T^2k^2e^{i\vec{k}\cdot(\vec{x}-\vec{x}^\prime)}e^{-i\mathcal{D}_Tk^2/2T}}{\sin(\frac{\mathcal{D}_Tk^2}{2T})(V/V^\prime)^{\frac{\mathcal{D}_Tk^2}{2\pi T}}}.
\end{align}
by choosing to close the contour down we pick up a pole at $\omega=-i\mathcal{D}_Tk^2$ and considering the near-horizon limit where $U\rightarrow0$.\\
\indent The bulk-to-boundary correlators are obtained by taking the limit $r^*\rightarrow0$. In this case, we have
\begin{align}
    K_{tx,tx}^{\text{ret.}}(\textbf{x};t_1,\vec{x}_1)&=-\int\frac{d^3k}{(2\pi)^3}\frac{\mathcal{D}_T^2k^2e^{i\vec{k}\cdot(\vec{x}-\vec{x}_1)}}{(Ve^{-2\pi Tt_1})^{\frac{\mathcal{D}_Tk^2}{2\pi T}}}\\&\;\;\;\;\;\;\;\;\;\;\times\Theta(V-e^{2\pi Tt_1})\nonumber,\\\nonumber
    &=\Theta(V-V_1)2\pi T\mathcal{D}_T\\&\;\;\; \times V\partial_V\int\frac{d^3k}{(2\pi)^3}\frac{e^{i\vec{k}\cdot(\vec{x}-\vec{x}^\prime)}}{(V/V_1)^{\frac{\mathcal{D}_Tk^2}{2\pi T}}},
\end{align}
for the retarded propagator and
\begin{align}
    K_{tx,tx}^{W}(\textbf{x}^\prime;t_1,\vec{x}_1)&=-\frac{i}{4}\int\frac{d^3k}{(2\pi)^3}\frac{\mathcal{D}_T^2k^2e^{i\vec{k}\cdot(\vec{x}^\prime-\vec{x}_1)}}{\sin(\frac{\mathcal{D}_Tk^2}{2T})}\\&\nonumber \;\;\; \; \times\frac{e^{-i\mathcal{D}_Tk^2/2T}}{(V^\prime e^{-2\pi Tt_1})^{\frac{\mathcal{D}_Tk^2}{2\pi T}}}\Theta(V^\prime-e^{2\pi Tt_1})\\\nonumber
    &=\frac{i\pi T\mathcal{D}_T}{2}\Theta(V^\prime-e^{2\pi Tt_1})V^\prime\partial_{V^\prime}\\&\nonumber \;\;\; \times\int\frac{d^3k}{(2\pi)^3}\frac{e^{i\vec{k}\cdot(\vec{x}^\prime-\vec{x}_1)}}{\sin(\frac{\mathcal{D}_Tk^2}{2T})}\frac{e^{-i\mathcal{D}_Tk^2/2T}}{(V^\prime/V_1)^{\frac{\mathcal{D}_Tk^2}{2\pi T}}},
\end{align}
for the Wightman propagator. Note that the dependence on $V_1$ cancels after multiplying the retarded and the Wightman bulk-to-boundary propagators. This means that there is no dependence on the insertion time $t_0$ of the first-order Green's function $G_{\mu\nu.\alpha\beta}^{(1)}(\textbf{x};\textbf{x}^\prime)$. However, as mentioned in \cite{Ahn}, the result should indeed depends on the insertion time. To overcome this problem, we use the perscription used in \cite{Cheng_2021}, so that the propagators become
\begin{align}
    K_{tx,tx}^{\text{ret.}}(\textbf{x};t_1,\vec{x}_1)&=\Theta(V-V_1)2\pi T\mathcal{D}_TV\\&\nonumber\;\;\;\; \times\partial_V\int\frac{d^3k}{(2\pi)^3}\frac{e^{i\vec{k}\cdot(\vec{x}-\vec{x}_1)}}{(V/V_1-1)^{\frac{\mathcal{D}_Tk^2}{2\pi T}}},\\\nonumber
    &=-\Theta(V-V_1)\frac{V}{V_1}\\&\nonumber\;\;\;\;\times \int\frac{d^3k}{(2\pi)^3}\frac{\mathcal{D}_T^2k^2e^{i\vec{k}\cdot(\vec{x}
    -\vec{x}_1)}}{(V/V_1-1)^{\frac{\mathcal{D}_Tk^2}{2\pi T}+1}},
\end{align}
and
\begin{align}
    K_{tx,tx}^{W}(\textbf{x};t_1,\vec{x}_1)&=\frac{i\pi T\mathcal{D}_T}{2}\Theta(V-V_1)V\partial_{V}\\&\nonumber \;\;\; \times\int\frac{d^3k}{(2\pi)^3}\frac{e^{i\vec{k}\cdot(\vec{x}^\prime-\vec{x}_1)}}{\sin(\frac{\mathcal{D}_Tk^2}{2T})} \frac{e^{-i\mathcal{D}_Tk^2/2T}}{(V/V_1-1)^{\frac{\mathcal{D}_Tk^2}{2\pi T}}}\\&\nonumber
    =\frac{i}{4}\Theta(V-V_1)\frac{V}{V_1}\int\frac{d^3k}{(2\pi)^3}\\&\nonumber \;\;\;\; \times\frac{\mathcal{D}_T^2k^2}{\sin(\frac{\mathcal{D}_Tk^2}{2T})} \frac{e^{i\vec{k}\cdot(\vec{x}-\vec{x}_1)}}{(1-V/V_1)^{\frac{\mathcal{D}_Tk^2}{2T}+1}}.
\end{align}
By taking $t_1\rightarrow-t_1-i\beta/2$ for the Wightman propagator, we get
\begin{align}
    K_{tx,tx}^W(\textbf{x};-t_1-i\beta/2&,\vec{x}_1)=-\frac{i}{4}VV_1\int\frac{d^3k}{(2\pi)^3}\\&\nonumber \;\;\; \times\frac{\mathcal{D}_T^2k^2}{\sin(\frac{\mathcal{D}_Tk^2}{2T})}\frac{e^{i\vec{k}\cdot(\vec{x}-\vec{x}_1)}}{(1+VV_1)^{\frac{\mathcal{D}_Tk^2}{2T}+1}}.
\end{align}
These propagators have qualitatively similar behaviors at late time limit ($V/V_1\gg1$) while maintaining the desired analytical properties connecting the retarded and Wightman propagators \cite{Cheng_2021}.\\
\indent The other propagators such as $K_{tx,zx}$ and $K_{zx,zx}$ can also be obtained from those derivations. Here, we have
\begin{align}
    K_{tx,zx}^{\text{ret.}}(\textbf{x};t_1,\vec{x}_1)\approx&-\Theta(V-V_1)\frac{V}{V_1}\\&\nonumber\times\int\frac{d^3k}{(2\pi)^3}\frac{i\mathcal{D}_T^3k^3e^{i\vec{k}\cdot(\vec{x}-\vec{x}_1)}}{(V/V_1-1)^{\frac{\mathcal{D}_Tk^2}{2\pi T}+1}},\\
    K_{zx,zx}^{\text{ret.}}(\textbf{x};t_1,\vec{x}_1)\approx&-\Theta(V-V_1)\frac{V}{V_1}\\&\nonumber\times\int\frac{d^3k}{(2\pi)^3}\frac{-\mathcal{D}_T^4k^4e^{i\vec{k}\cdot(\vec{x}-\vec{x}_1)}}{(V/V_1-1)^{\frac{\mathcal{D}_Tk^2}{2\pi T}+1}},
\end{align}
and
\begin{align}
    K_{tx,zx}^W(&\textbf{x};-t_1-i\beta/2,\vec{x}_1)\approx-\frac{i}{4}VV_1\\&\nonumber\;\; \times\int\frac{d^3k}{(2\pi)^3}\frac{i\mathcal{D}_T^3k^3}{\sin(\frac{\mathcal{D}_Tk^2}{2T})}\frac{e^{i\vec{k}\cdot(\vec{x}-\vec{x}_1)}}{(1+VV_1)^{\frac{\mathcal{D}_Tk^2}{2T}+1}},\\K_{zx,zx}^W(&\textbf{x};-t_1-i\beta/2,\vec{x}_1)\approx-\frac{i}{4}VV_1\\&\nonumber\;\; \times\int\frac{d^3k}{(2\pi)^3}\frac{-\mathcal{D}_T^4k^4}{\sin(\frac{\mathcal{D}_Tk^2}{2T})}\frac{e^{i\vec{k}\cdot(\vec{x}-\vec{x}_1)}}{(1+VV_1)^{\frac{\mathcal{D}_Tk^2}{2T}+1}}.
\end{align}
The main difference between the three cases are the order of $k$ inside the bulk-to-boundary propagators. This will affect the traversability of the wormhole since higher order in $k$ could lead to lower traversability in the hydrodynamic limit.
\begin{figure*}
    \centering
\includegraphics[width=0.8\columnwidth]{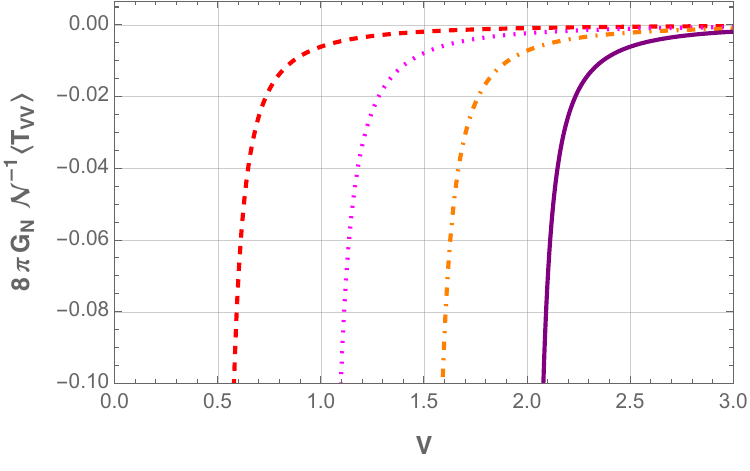}
\includegraphics[width=0.8\columnwidth]{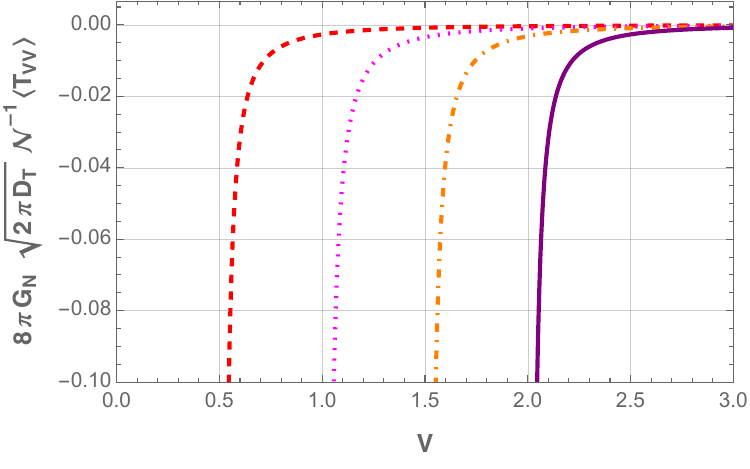}\\
\includegraphics[width=1\columnwidth]{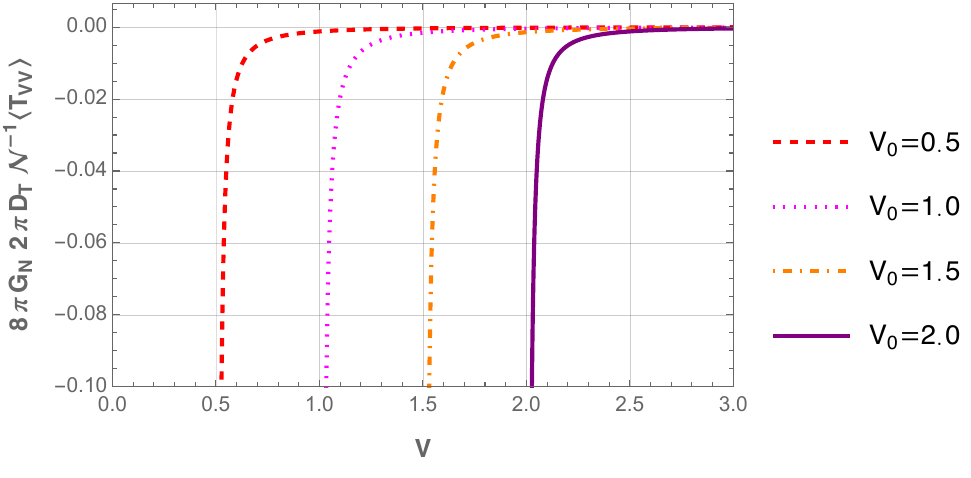}
\caption{\label{fig:case1shear}Normalized energy-momentum tensor versus $V$ for various values of the insertion time $V_0$ for Case I (top-left), Case II (top-right), and Case III (bottom). Here, we use $\bar{a}=-1$ and $u_{\text{max}}=1/2$.}
    \end{figure*}
\subsubsection{Violation of the ANEC}
After calculating the bulk-to-boundary propagators, we are now ready to see the violation of the ANEC. The first-order Green's function for Case I is then given by
\begin{widetext}
\vspace{-1em}
    \begin{align}
    G_{tx,tx}^{(1)}(\textbf{x};\textbf{x}^\prime)&=i\int_{t_0}^{t^\prime} dt_1d^3x_1A(t_1)K_{tx,tx}^{W}(\textbf{x}^\prime;-t_1-i\beta/2,\vec{x}_1)K_{tx,tx}^{\text{ret.}}(\textbf{x};t_1,\vec{x}_1)+\textbf{x}\leftrightarrow\textbf{x}^\prime\\\nonumber
    &=-\frac{VV^\prime}{8\pi T}\int_{V_0}^{V^\prime}\frac{dV_1}{V_1}A(V_1)\int\frac{d^3k}{(2\pi)^3}\frac{\mathcal{D}_T^4}{\sin(\frac{\mathcal{D}_Tk^2}{2T})}\frac{k^4e^{i\vec{k}\cdot(\vec{x}^\prime-\vec{x})}}{[(V/V_0-1)(1+V^\prime V_0)]^{\frac{\mathcal{D}_Tk^2}{2\pi T}+1}}+\textbf{x}\leftrightarrow\textbf{x}^\prime.
\end{align}
By choosing a delta function perturbation inserted at $t=t_0$ or $A(V_1)=aV_0\delta(V_1-V_0)$, where $a$ is a constant, the expectation value of $\hat{T}_{VV}$ in Eq. \eqref{ANECexpval} is then given by
\begin{align}
    8\pi G_{\text{N}}\langle\hat{T}_{VV}^{(2)}\rangle&=\frac{a\mathcal{D}_T^4}{32\pi^3T^3}\frac{1}{P(V,V_0)}\int\frac{d^3k}{(2\pi)^3}\frac{1}{4\sin(\frac{\mathcal{D}_Tk^2}{2T})}\frac{k^6}{P(V,V_0)^{\frac{\mathcal{D}_Tk^2}{2\pi T}}}\\\nonumber
    &=\frac{a\mathcal{D}_T^4}{32\pi^3T^3}\frac{1}{P(V,V_0)}\int_0^{k_{\text{max}}}\frac{dk}{(8\pi^2)}\nonumber\frac{1}{\sin(\frac{\mathcal{D}_Tk^2}{2T})}\frac{k^8}{P(V,V_0)^{\frac{\mathcal{D}_Tk^2}{2\pi T}}},
\end{align}
\end{widetext}
or we can write it as
\begin{align}
    8\pi G_{\text{N}}\langle&\hat{T}_{VV}^{(2)}\rangle=\frac{a\mathcal{D}_T^4}{2^9\pi^{\frac{9}{2}}\Gamma(\frac{3}{2})T^3}\frac{1}{P(V,V_0)}\\&\nonumber\;\;\;\times\int_0^{k_{max}}dk\frac{1}{\sin(\frac{\mathcal{D}_Tk^2}{2T})}\frac{k^8}{{P(V,V_0)^{\frac{\mathcal{D}_Tk^2}{2\pi T}}}},
\end{align}
where
\begin{equation}
    P(V,V_0)\equiv(V/V_0-1)(1+VV_0).
\end{equation}
The expression of $\langle\hat
{T}_{VV}^{(2)}\rangle$ is similar to the result for the vector perturbation in \cite{Ahn}, but differs by $2^{-5}$ which comes from a factor present in the $\langle \hat{T}_{VV}^{(2)}\rangle$ component of the equation \eqref{Tvv}. 
Since we are only working within the hydrodynamic limit where $\frac{k}{2\pi T}\ll1$, therefore, we need to impose a hard cut-off at $k_{\text{max}}\approx 2\pi T\sqrt{f}$ \cite{Ahn}, where $f$ is a parameter that controls the maximum energy of the deformation.\\
\begin{figure}
    \centering
\includegraphics[width=0.9\linewidth]{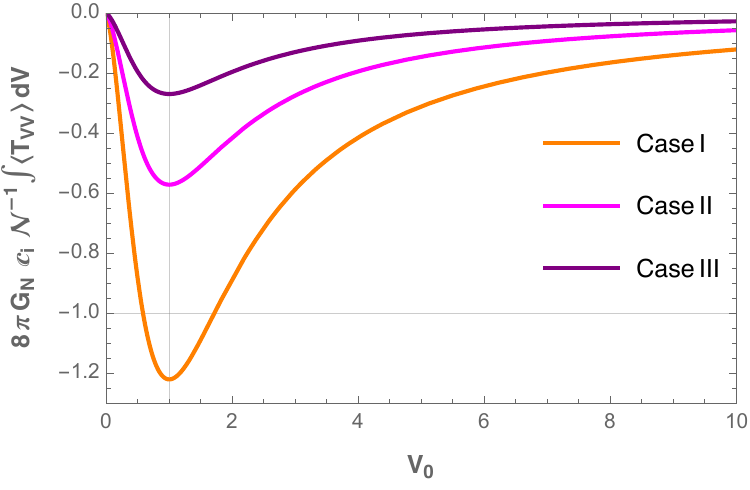}
\caption{\label{fig:case1shearANECintegral} The normalized ANEC as a function of insertion time $V_0$ for Case I, II, and III that ends at $V=+\infty$. Here, we use $\bar{a}=-1$ and $u_{\text{max}}=1/2$. We can see, the earlier the deformation is turned on the more traversable the wormhole shown by the increasing negative value of ANEC. Note that the insertion time $V_0=1$ corresponds to insertion time $t_0=0$, where the wormhole becomes most traversable.}
\end{figure}
\indent The integration can be performed numerically by first defining a parameter $u=\frac{\mathcal{D}_Tk^2}{2\pi T}$, so that the energy-momentum tensor becomes
\begin{align}\label{ANECresultvector}
    8\pi &G_{\text{N}}\langle\hat{T}_{VV}^{(2)}\rangle=\frac{a\mathcal{D}_T^4}{2^{10}\pi^{\frac{9}{2}}\Gamma(\frac{3}{2})}\bigg(\frac{2\pi T}{\mathcal{D}_T}\bigg)^{\frac{1}{2}}\bigg(\frac{2\pi T}{\mathcal{D}_T}\bigg)^{4}\\\nonumber&\;\;\;\times\frac{1}{P(V,V_0)}\int_0^{u_{\text{max}}}\frac{u^\frac{7}{2}}{\sin(\pi u)P(V,V_0)^u}du\\\nonumber
    &=\frac{\bar{a}}{2^5\sqrt{2}\Gamma(\frac{3}{2}) \bar{\mathcal{D}}_T^{\frac{1}{2}}}\frac{1}{P(V,V_0)}\int_0^{u_{\text{max}}}\frac{u^\frac{7}{2}}{\sin(\pi u)P(V,V_0)^u}du,
\end{align}
where $u_{\text{max.}}=2\pi\bar{\mathcal{D}}_T f$. Here, we rescale the parameters $\mathcal{D}_T\rightarrow\bar{\mathcal{D}}_TT^{-1}$ and $a\rightarrow \bar{a}T^{-2}$ to introduce dimensionless parameters $\bar{\mathcal{D}}_T$ and $\bar{a}$. This result reproduce the function obtained in \cite{Ahn} for double trace deformation from $U(1)$ gauge field up to a proportionality dimensionless constant. We then define the normalization constant $\mathcal{N}(\bar{\mathcal{D}}_T)=\frac{1}{2^5\sqrt{2}\Gamma(\frac{3}{2})\bar{\mathcal{D}}_T^{\frac{1}{2}}}$.\\
\indent To see the traversability of the wormhole, we plot the normalized energy-momentum tensor and the normalized ANEC relatives to insertion time $V_0$ in Figures \ref{fig:case1shear} and \ref{fig:case1shearANECintegral}, respectively. The figures show that the wormhole is traversable, given a negative value of $\bar{a}$ which indicates negative-energy shock waves. In this work, we extend the vector mode calculations to more general cases involving Case II and Case III as well. We obtain the $\langle\hat{T}_{VV}^{(2)}\rangle$ for each case, which is given by
\begin{equation}
\langle\hat{T}_{VV}^{(2)}\rangle=\frac{\mathbb{c}_i\mathcal{N}(\bar{\mathcal{D}}_T)}{P(V,V_0)}\int_0^{u_{\text{max.}}}\frac{u^{m_i}}{\sin(\pi u)P(V,V_0)^u}du,
\end{equation}
where $\mathbb{c}_i$ are given by
\begin{align}
    \mathbb{c}_{\text{\tiny I}}=1,\;\;\;\mathbb{c}_{\text{\tiny II}}=\sqrt{2\pi \bar{\mathcal{D}}_T},\;\;\;\mathbb{c}_{\text{\tiny III}}=2\pi\bar{\mathcal{D}}_T,
\end{align}
and $m_i$ are given by
\begin{align}
    m_{\text{\tiny I}}=\frac{7}{2},\;\;\;m_{\text{\tiny II}}=\frac{8}{2},\;\;\;m_{\text{\tiny III}}=\frac{9}{2},
\end{align}\\
for Case I, II, and III, respectively.
\subsection{Sound Channel}
\begin{figure*}
    \centering
    \includegraphics[scale=0.57]{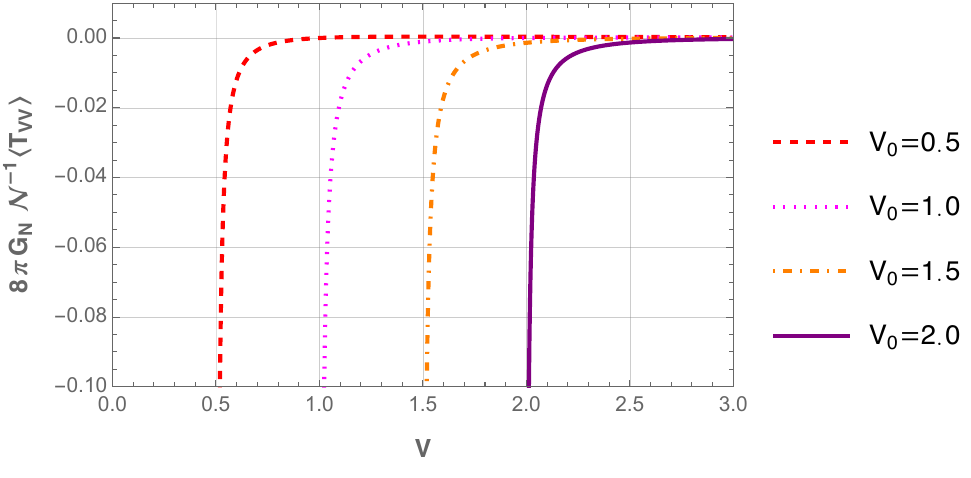}\;\;\;
    \includegraphics[scale=0.57]{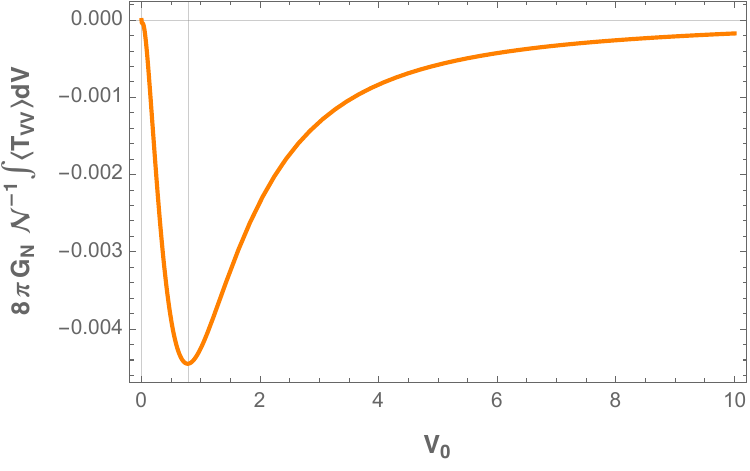}
    \caption{The plot of the normalized $\langle\hat{T}_{VV}^{(2)}\rangle$ versus $V$ (left) and the ANEC versus $V_0$ (right) for the sound channels with $u_{\text{max.}}=1/2$.}
\label{fig:soundmodesanecplot}
\end{figure*}
\begin{figure*}
    \centering
    \includegraphics[width=0.58\linewidth]{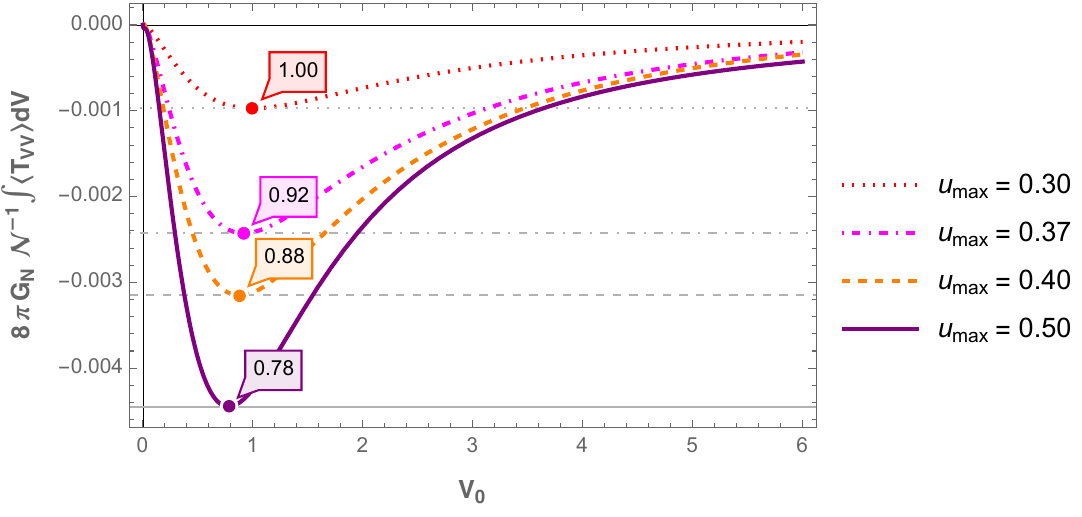}
    \caption{Plot of ANEC versus $V_0$ for various values of $u_{\text{max.}}$ in sound channels. The markers indicate the values of $V_0$ corresponding to the minimum ANEC for each $u_{\text{max.}}$.}
    \label{fig:variasiumax}
\end{figure*}
In this part, we calculate the ANEC from double trace deformation due to the sound channel perturbations. In the sound channel, the nonvanishing gravitational perturbation components are
\begin{equation}
\mathfrak{h}_{tt},\;\mathfrak{h}_{tz},\;\mathfrak{h}_{xx}=\mathfrak{h}_{yy},\;\mathfrak{h}_{zz}.
\end{equation}
Therefore, the metric in the Kruskal coordinates is given by
\begin{widetext}
\begin{align}\label{perturbedmetricsound}
    ds^2=&-\frac{\beta^2}{4\pi^2 UV}g_{tt}(UV)dUdV+g_{xx}(UV)(dx^2+dy^2+dz^2)+\frac{\beta^2}{16\pi^2U^2V^2}g_{xx}(UV)\mathfrak{h}_{tt}(U,V,z)(UdV-VdU)^2\\
    &+g_{xx}(UV)\mathfrak{h}_{xx}(U,V,z)(dx^2+dy^2)+g_{xx}(UV)\mathfrak{h}_{zz}(U,V,z)dz^2+\frac{\beta}{2\pi UV}g_{xx}(UV)\mathfrak{h}_{zt}(U,V,z)(UdV-VdU)dz.\nonumber
\end{align}
\end{widetext}
In this metric, the energy-momentum tensor that is quadratic in $h_{MN}$ near the horizon where $U\rightarrow 0$ is given by (Appendix A)
\begin{align}\label{Tvvsound}
    \langle\hat{T}_{VV}^{(2)}\rangle=-&\frac{1}{8\pi G_{\text{N}}}\frac{\beta^2}{64\pi^2 V^2}\big(\langle\partial_z\mathfrak{h}_{tt}\partial_z\mathfrak{h}_{tt}\rangle+2\langle\mathfrak{h}_{tt}\partial_z^2\mathfrak{h}_{tt}\rangle\big)\\\nonumber
    & +\mathcal{O}(\beta).
\end{align}
Using the point-splitting method, the energy momentum tensor now becomes
\begin{align}
    \langle\hat{T}_{VV}^{(2)}\rangle=&-\frac{1}{8\pi G_{N}}\frac{\beta^2}{64\pi^2 V^2}\\\nonumber
    &\times\lim_{\textbf{x}^\prime\rightarrow \textbf{x}}\big(\partial_z\partial_{z^\prime}+2\partial_{z^\prime}^2)\langle\mathfrak{h}_{tt}(\textbf{x})\mathfrak{h}_{tt}(\textbf{x}^\prime)\rangle.
\end{align}
This expression can be simplified even further. Note that the Green's functions depend on $z$ and $z^\prime$ through the factor $e^{ik(z-z^\prime)}$. Therefore, we may change $\partial_{z^\prime}\rightarrow-\partial_z$. In this expression, we have
\begin{align}
    \langle\hat{T}_{VV}^{(2)}\rangle
    =&\frac{1}{8\pi G_{\text{N}}}\frac{\beta^2}{64\pi^2 V^2}\lim_{\textbf{x}^\prime\rightarrow\textbf{x}}\partial_z\partial_{z^\prime}G_{tt,tt}^{(1)}(\textbf{x};\textbf{x}^\prime),
\end{align}
where $G_{\mu\nu,\alpha\beta}^{(1)}(\textbf{x};\textbf{x}^\prime)$ is the first-order Green's function given by Eq. \eqref{firstordergreen}. \\
\indent For now, we consider the simplest case where the coupling function of the double trace deformation is in the form of $Q^{\lambda\rho\gamma\sigma}(t,\vec{x})=A(t)\delta^\lambda_t\delta^\rho_t\delta^\gamma_t\delta^\sigma_t$ with a delta function coupling $A(t)=aV_0\delta(V-V_0)$ as in the previous cases. In this case, we need to calculate the bulk-to-boundary propagators $K^W_{tt,tt}(\textbf{x};t_1,\vec{x}_1)$ and $K_{tt,tt}^{\text{ret.}}(\textbf{x};t_1,\vec{x}_1)$ from the bulk-to-bulk propagators similar to the ones done for the shear channel. For the sound channel, the retarded bulk-to-bulk propagator in the Fourier space within the hydrodynamic limit is given by
\begin{equation}
G_{\mu\nu,\alpha\beta}^{\text{ret.}}(r^*_1,r_2^*,\omega,k)\approx\frac{\mathbb{S}_{\mu\nu,\alpha\beta}(\omega,k)e^{-i\omega(r_1^*-r_2^*)}}{(\omega^2-v_s^2k^2)T},
\end{equation}
where $v_s$ is the sound speed and $v_s=\frac{1}{\sqrt{3}}$ for Schwarzschild-AdS$_5$ solution. The factor $T$ is inserted in the denominator to ensure that the dimension is correct and agrees with the ones in shear channel. The function $\mathbb{S}_{\mu\nu,\alpha\beta}(\omega,k)$ for the sound channels are given by Table \ref{table1} \cite{MATSUO2009593}.
\begin{center}
\begin{table}[ht]
\caption{\label{table1}The value of $\mathbb{S}_{\mu\nu,\alpha\beta}(\omega,k)$ for various $\mu\nu$ and $\alpha\beta$ indices}
    \begin{tabular}{c||c|c|c|c|}
         & $tt$ & $xx$ & $zz$ & $zt$ \\
         \hline\hline
         $tt$ & $(5k^2-3\omega^2)$ & $2(k^2+\omega^2)$ & $(k^2+\omega^2)$ & $8(k^2+\omega^2)$  \\
         \hline
         $xx$ & - & $16\omega^2/3$ & $2(k^2+\omega^2)/3$ & $16k\omega/3$ \\
         \hline
         $zz$ & - & - & $(-k^2+7\omega^2)/3$ & $8k\omega/3$\\
         \hline
         $zt$ & - & - & - & $4(k^2+9\omega^2)/3$\\
         \hline
    \end{tabular}
    \end{table}
\end{center}
\par The sound channel Green's functions have two real poles located at $\omega=+v_sk$ and $\omega=-v_sk$. When evaluating the $\omega$ integration, we pick both poles that are located on the real axis. Furthermore, if we consider the attenuation constant term ($\sim-i\Gamma_sk^2$), the poles of the Green's functions are instead at $\omega=\pm v_sk-i\Gamma_s k^2$, placing them slightly below the real axis. In this case, by closing the contour downward, we pick up both poles, and the retarded prescription is automatically realized through the Heaviside step function $\Theta(V - V^\prime)$. Therefore, real-space retarded Green's function $G_{tt,tt}^{\text{ret.}}(\textbf{x};\textbf{x}^\prime)$ without the attenuation term can be expressed as
\begin{align}
    G^{ret}_{tt,tt}(\textbf{x};\textbf{x}^\prime)=&\frac{3iv_s}{T}\Theta(V-V_1)\int \frac{d^3k}{(2\pi)^3}k e^{ik(\textbf{x}-\textbf{x}^\prime)}\\\nonumber
    &\times\left[ \left(\frac{V}{V^\prime}\right)^{\frac{ikv_s}{2\pi T}}-\left(\frac{V}{V^\prime}\right)^{-\frac{ikv_s}{2\pi T}}\right].
\end{align}
Following the similar reasoning as the computations done for the shear channels, the retarded bulk-to-boundary propagator is given by
\begin{align}  K^{\text{ret.}}_{tt,tt}(\textbf{x};t_1,x_1)=&\frac{6v_s}{T}\Theta(V-V_1)\frac{V/V_1}{V/V_1-1}\\\nonumber
&\times\int \frac{d^3k}{(2\pi)^3}k e^{ik(\textbf{x}-\textbf{x}^\prime)}\Im(\mathcal{R}(V,V_1,k)),
\end{align}
where $V_1=e^{2\pi Tt_1}$ and $\mathcal{R}(V,V_1,k)$ is defined as
\begin{equation}
    \mathcal{R}(V,V_1,k)\equiv(V/V_1-1)^{-\frac{ikv_s}{2\pi T}}.
\end{equation}
The Wightman Green's function is again obtained from the relation in Eq. \eqref{Wretrelation} and hence the bulk-to-boundary propagator can be expressed as
\begin{align}
K_{tt,tt}^W(\textbf{x}&;-t_1-i\beta/2,\vec{x}_1)=-\frac{3iv_s}{2T}\frac{VV_1}{1+VV_1}\\\nonumber&\times\int\frac{d^3k}{(2\pi)^3}\frac{ke^{i\vec{k}(\vec{x}-\vec{x}_1)}}{\sinh\big(\frac{kv_s}{2T}\big)}\Re(\mathcal{Q}(V,V_1,k)),
\end{align}
where $\mathcal{Q}(V,V_1,k)$ is defined as
\begin{equation}
\mathcal{Q}\equiv(1+VV_1)^{-\frac{ikv_s}{2\pi T}}.
\end{equation}
\indent The first-order Green's function for the sound channel is now given by
\begin{widetext}
\begin{align}
    G_{tt,tt}^{(1)}(\textbf{x};\textbf{x}^\prime)&=i\int_{t_0}^t dt_1d^3x_1A(t_1)K_{tt,tt}^{W}(\textbf{x}^\prime;-t_1-i\beta/2,\vec{x}_1)K_{tt,tt}^{\text{ret.}}(\textbf{x};t_1,\vec{x}_1)+\textbf{x}\leftrightarrow\textbf{x}^\prime\\\nonumber
    =&\frac{3aVV^\prime/(2\pi T^3)}{(V/V_0-1)(1+V^\prime V_0)}\int\frac{d^3k}{(2\pi)^3}\frac{k^2e^{i\vec{k}\cdot(\vec{x}-\vec{x}^\prime)}}{\sinh\big(\frac{kv_s}{2T}\big)}\Re(\mathcal{Q}(V^\prime,V_0,k))\Im(\mathcal{R}(V,V_0,k))+\textbf{x}\leftrightarrow\textbf{x}^\prime.
\end{align}
From this, the expectation value of $\hat{T}_{VV}$ can now be expressed as
\begin{align}
    8\pi G_{\text{N}}\langle\hat{T}_{VV}\rangle=&\frac{a\beta^2/(2\pi T^3)}{64\pi^2P(V,V_0)}\int\frac{d^3k}{(2\pi)^3}\frac{k^4}{\sinh\big(\frac{kv_s}{2T}\big)}\Re(\mathcal{Q}(V,V_0,k))\Im(\mathcal{R}(V,V_0,k))\\\nonumber
    =&\frac{a\beta^2}{2^8\pi^5T^3P(V,V_0)}\int_0^{k_{\text{max.}}} dk\frac{k^6}{\sinh\big(\frac{kv_s}{2T}\big)}\Re(\mathcal{Q}(V,V_0,k))\Im(\mathcal{R}(V,V_0,k)),
\end{align}
\end{widetext}
where we impose a hard cut-off at $k_{\text{max.}}=2\pi T f_s$, with $f_s$ is an $\mathcal{O}(1)$ constant analogous to the shear channel one. By defining a dimensionless parameter $u\equiv\frac{kv_s}{2\pi T}$, we have
\begin{align}
    8\pi G_{\text{N}}\langle&\hat{T}_{VV}\rangle=\frac{3^{7/2}\pi^2\bar{a}}{P(V,V_0)}\int_0^{u_{\text{max.}}} du\frac{u^6}{\sinh(\pi u)}\\&\nonumber\;\;\times\Re(\mathcal{Q}(V,V_0,k(u)))\Im(\mathcal{R}(V,V_0,k(u))).
\end{align}
The energy-momentum tensor can now be written as
\begin{align}
    8\pi G_{\text{N}}\langle&\hat T_{VV}\rangle=\frac{\mathcal{N}\bar{a}}{P(V,V_0)}\int_0^{u_{\text{max.}}} du\frac{u^6}{\sinh(\pi u)}\\&\nonumber\;\;\times\Re(\mathcal{Q}(V,V_0,k(u)))\Im(\mathcal{R}(V,V_0,k(u))),
\end{align}
where
\begin{equation}
\mathcal{N}=3^{7/2}\pi^2,
\end{equation}
is the normalization constant for sound channel perturbations. Note that this normalization constant only depends on the dimensionless sound speed $v_s=\frac{1}{\sqrt{3}}$. Once we generalize the speed of sound, $u_{\text{max.}}$ now becomes $u_{\text{max.}}=v_s f_s$. Therefore, one could expect that models with lower speed also lead to narrower wormhole opening.\\
\indent The plot of the normalized energy–momentum tensor as a function of the coordinate $V$ and the ANEC as a function of the insertion time $V_0$ for the sound channels is shown in Figure \ref{fig:soundmodesanecplot}. It can be seen that the wormhole is also traversable in this case. However, the degree of wormhole traversability is lower than in the shear channels, as indicated by the smaller absolute value of the ANEC. After varying the value of the sound speed (via variation of $u_{\text{max.}}$), the result can be seen in Figure \ref{fig:variasiumax}. It is shown that higher $v_s$ leads to broader wormhole opening. We also find that, with the propagating sound channels, the value of $V_0$ that leads to the minimum ANEC is also influenced by the speed of sound. Higher $v_s$ requires earlier insertion time for the double trace deformation. Moreover, the sound channels also shows a power-law remnant at late times, which will be presented in more detail in the next section using a fitting model.\\
\begin{figure*}
    \centering
\includegraphics[width=0.6\linewidth]{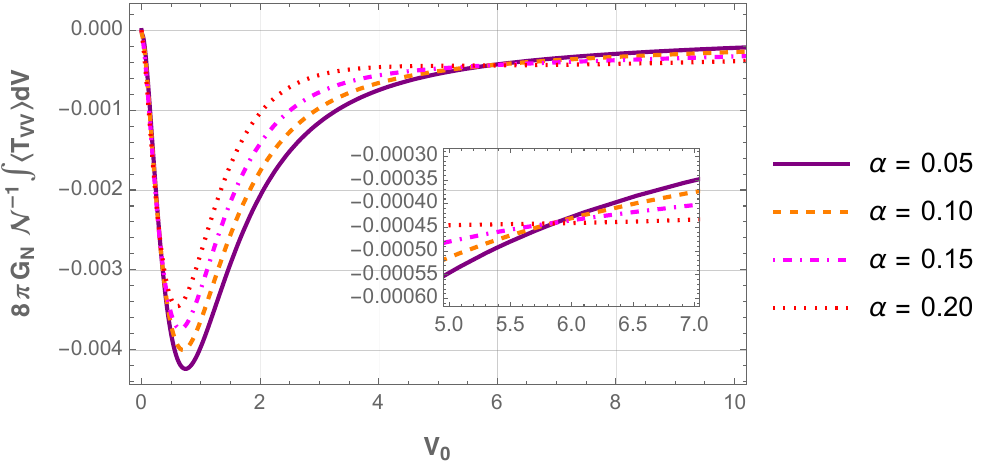}
    \caption{The plot of ANEC versus $V_0$ for various values of the ratio $\alpha\equiv\frac{2\pi \bar{\Gamma}_s}{v_s^2}$ in sound channel. In this plot, we choose $u_{\text{max.}}=1/2$.} 
    \label{fig:varyattenuation}
\end{figure*}
\indent For sound channels, the speed of sound is not the only important parameter, as the attenuation constant $\Gamma_s$ also plays a crucial role, which we will reveal soon. If we include the damping or the attenuation of the sound channels, with the pole of the frequency is now given by $\omega\sim\pm v_sk -i\Gamma_s k^2$, the energy-momentum tensor is now given by
\begin{align}\label{TVVattenuation}
    8\pi G_{\text{N}}\langle\hat{T}_{VV}&\rangle=\frac{\mathcal{N}\bar{a}}{P(V,V_0)}\int_0^{u_{\text{max.}}} du\frac{u^6}{\sinh(\pi u)}\\\nonumber
    &\times\frac{\Re(\mathcal{Q}(V,V_0,k(u)))\Im(\mathcal{R}(V,V_0,k(u)))}{P(V,V_0)^{\frac{2\pi \bar{\Gamma}_su^2}{v_s^2}}},
\end{align}
where $\bar{\Gamma}_s=\Gamma_s T$ is the dimensionless attenuation constant. The ratio $\frac{\bar{\Gamma}_s}{v_s^2}$ now plays a role in shaping the ANEC function. We still use the value $u_{\text{max.}}=v_s f$.\\
\indent Here, we vary the value of the attenuation constant by varying $\alpha\equiv\frac{2\pi \bar{\Gamma}_s}{v_s^2}$. The result can be seen in Figure \ref{fig:varyattenuation}. We find that increasing the attenuation constant suppresses traversability at smaller $V_0$, but enhances it at larger $V_0$. Moreover, varying $\alpha$ also changes the insertion time $V_0$ corresponding to the minimum ANEC. This happens because $\alpha$ depends on both the dimensionless attenuation constant $\bar{\Gamma}_s$ and the sound speed $v_s$, whose effects compete with each other for different values of $V_0$.\\
\indent In sound channels, one could also study the traversability of the wormhole when the speed of sound reaches the speed of light, or even surpass it, leading to superluminal gravitational perturbations. In Figure \ref{fig:superluminal}, one can see that from $v_s=v_{\text{Schw.}}$ up until $v_s=1$, increasing speed leads to increasing traversability. However, as the speed of sound becomes greater than the speed of light, the traversability gets smaller. There, the wormhole opens up right after $t = t_0$ and subsequently closes again shortly thereafter, as a larger $t_0$ leads to shorter opening time. For sufficiently late insertion times, the opening becomes too transient for a signal or particle to traverse before the wormhole closes, indicated by a positive value of ANEC. It can be seen in Figure \ref{fig:superluminal} (bottom right) when the insertion time is chosen such that $V_0=10$. This behavior may arise because the superluminal modes induce tachyonic probes that might counteract the effect of the negative-energy shock waves.
\begin{figure*}
    \centering
    \includegraphics[width=0.65\linewidth]{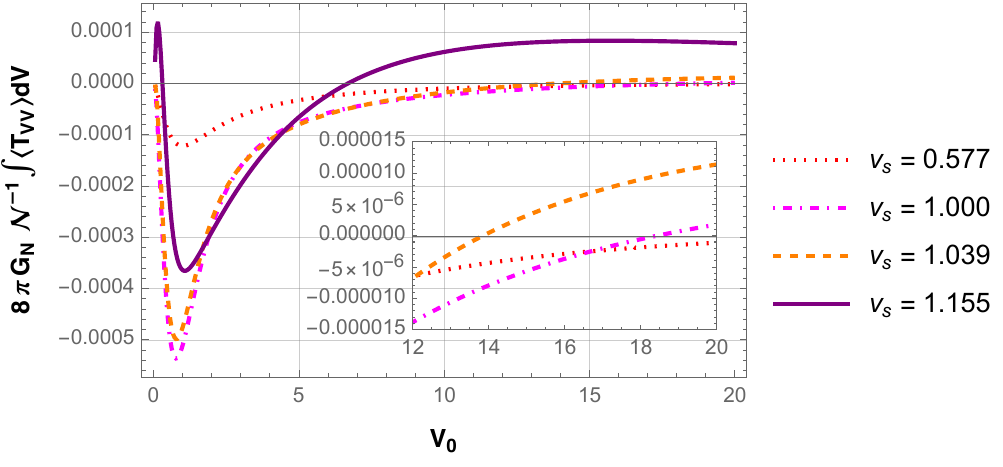}\\   \includegraphics[width=0.45\linewidth]{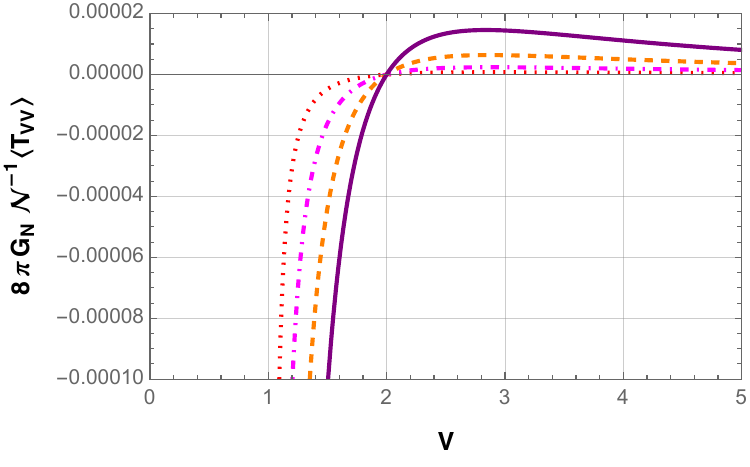}\;
\includegraphics[width=0.45\linewidth]{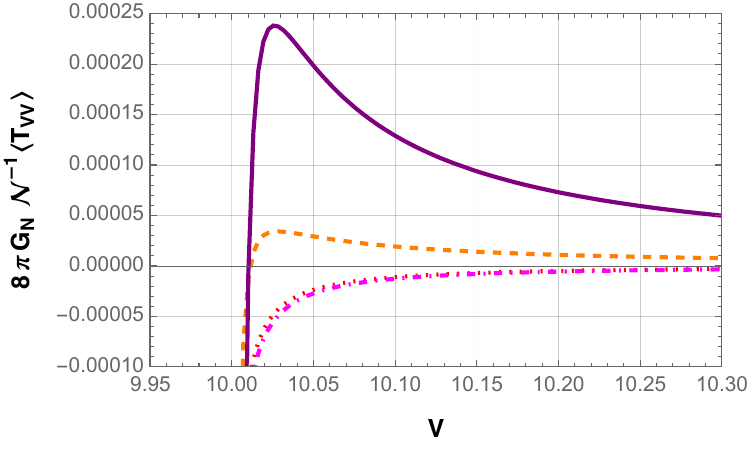}
    \caption{The plot of ANEC versus $V_0$ (top) and the plot of NEC versus $V$ at $V_0=1$ (bottom-left) and at $V_0=10$ (bottom-right) for various values of the speed of sound $v_s$ where the speed of light is set to $c=1$. Here we choose the attenuation parameter as $2\pi \bar{\Gamma}_s=0.002$ and $f_s=1/2$.}
    \label{fig:superluminal}
\end{figure*}
\section{Data Fitting and Power-law Remnants}\label{section4}
\begin{figure}
    \centering
\includegraphics[width=1\linewidth]{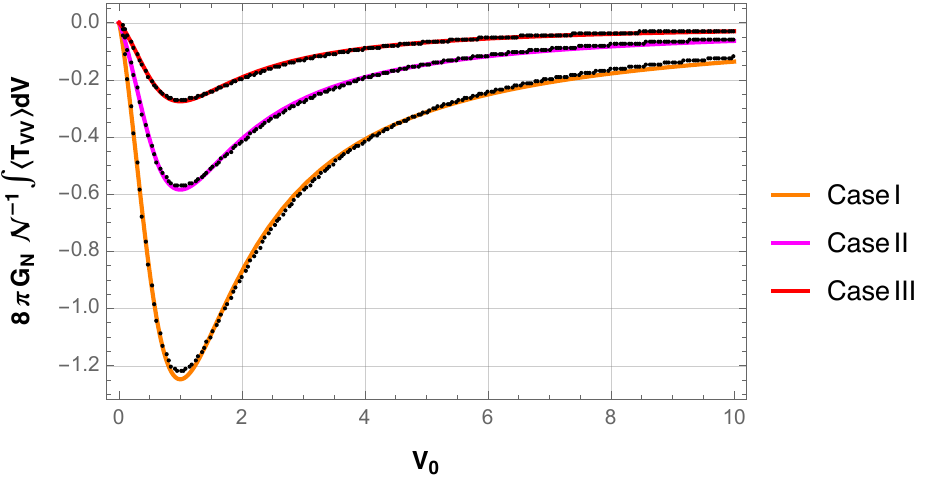}
    \caption{Comparison between the new fitting model (solid lines) and the actual result from Eq. \eqref{ANECresultvector} (dots) of the normalized ANEC as a function of $V_0$ for shear channel.}
    \label{fig:fitvector}
\end{figure}

\begin{figure*}
    \centering
    \includegraphics[scale=0.58]{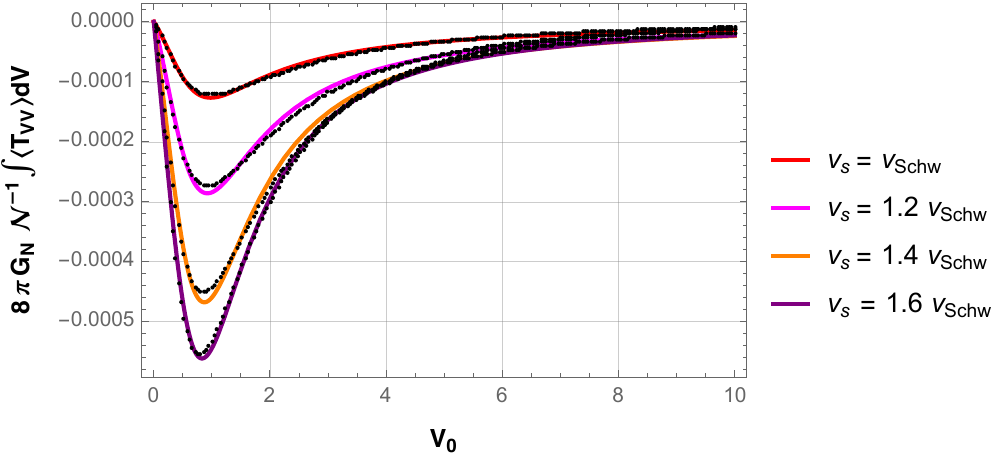}\;\;
    \includegraphics[scale=0.55]{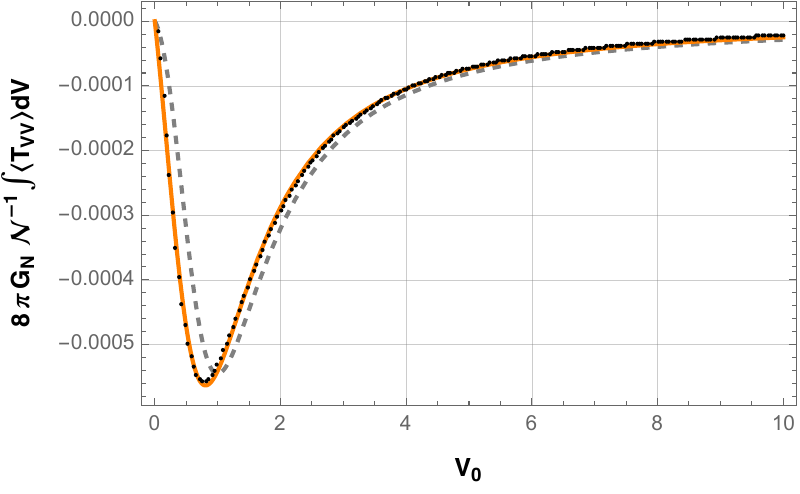}
    \caption{Left: Results from Eq. \eqref{TVVattenuation} (dots) and the new fitting model (solid lines) of the ANEC when $v_s$ is varied. Right: Comparison between the old fitting model (gray-dashed), new fitting model (orange-solid), and the result from Eq. \eqref{TVVattenuation} (dots) of the normalized ANEC as a function of $V_0$ for sound channel. Here, we fix $2\pi \bar{\Gamma}_s=0.002$ and $f_s=1/2$.}
    \label{fig:fitsound}
\end{figure*}

\begin{figure}
    \centering
\includegraphics[width=0.9\linewidth]{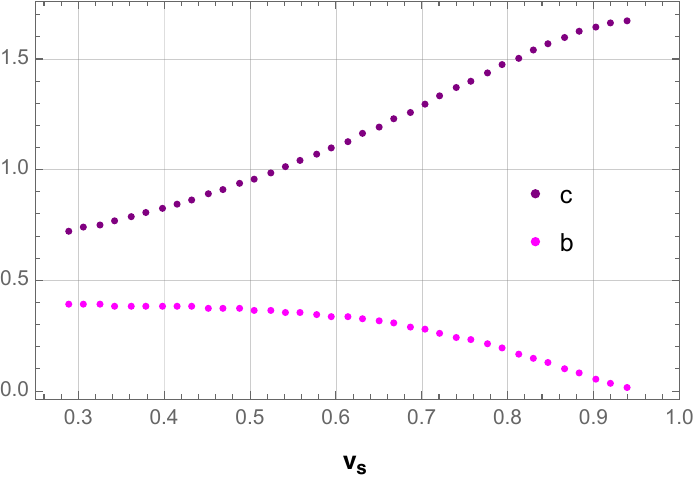}
    \caption{The values of the fitting parameters $c$ and $b$ for various sound speed $v_s$ ranging from $v_s=0.5v_{\text{Schw.}}$ to $v_s\approx0.95$.}
    \label{fig:canbfit}
\end{figure}

In this section, we consider a fitting model that can describe the ANEC for both shear and sound channels. It has already shown in \cite{Ahn} that the $U(1)$ gauge field (which is comparable to the shear channel in our case) can be well described by a fitting model
\begin{equation}
    \mathcal{A}(V_0)=\int_{V_0}^\infty \langle\hat{T}_{VV}^{(2)}\rangle dV\approx-\frac{V_0}{V_0^2+1}\frac{a}{[\ln(V_0+1/V_0)]^b}.
\end{equation}
This function exhibits a power-law prefactor in the  limit of sufficiently large $ t_0 $, under the condition $V_0 \gg 1$,
\begin{equation}\label{av0ahn}
\mathcal{A}(V_0)\approx -\frac{ae^{-2\pi T t_0}}{t_0^b}.
\end{equation}
\indent The value of the exponent $b$ can be obtained by fitting the model to the result from Eq. \eqref{ANECresultvector}.
In this work, we slightly generalize the fitting model to describe both the gravitational shear and sound channels simultaneously, achieving a better overall match. We consider ANEC in the form of
\begin{equation}
    \mathcal{A}_{\text{new}}(V_0)=-\frac{V_0}{V_0^{(c+1)}+1}\frac{a}{[\ln(V_0+1/V_0)]^b},
\end{equation}
where now $\{a,b,c\}$ are the fitting parameters and the values $c=1$ correspond to the old model. We find that, for the shear channels, the values of the fitting parameters are given by Table \ref{table2} and the plot of the fitting function can be seen in Figure \ref{fig:fitvector}. As in Eq. \eqref{av0ahn}, the new model retains a power law factor at sufficiently large times, leading to the form
\begin{equation}\label{powerlawtail}
    \mathcal{A}_{\text{new}}(V_0)\sim-\frac{e^{-c2\pi Tt_0}}{t_0^b},
\end{equation}
where the exponential decay rate now acquires an additional dependence on parameter $c$. For sufficiently large, yet still intermediate, timescales, the power law factor can provide a noticeable correction to the exponential envelope. However, in all cases involving scalar, vector, and tensor operators, the exponential part dominates the late-time regime. The power-law factor also appear in the two-point function of the stress tensor correlators \cite{Kovtun2003hydrodynamics,CaronHuot2010} and in the out-of-time ordered correlator (OTOC) of stress-energy tensor \footnote{please refer to Appendix B of \cite{Cheng_2021}}.\\
\indent Both parameters \(b\) and \(c\) govern whether the solution exhibits a purely exponential behavior or, alternatively, an exponential behavior modified by an additional power-law factor at late times. In the shear channels, the result remains largely unchanged, whereas in the sound channels, we observe noticeable differences.
\begin{table}
\centering
\caption{\label{table2}The values of the fitting parameters $\{a,b,c\}$ for both Old and New models in shear channels.}
\vspace{2mm}
\begin{tabular}{|p{2cm}|p{1.5cm}|p{1.5cm}|p{1.5cm}|p{1.5cm}|}
    \hline
    Old Model & $a$ & $b$ & $c$\\
    \hline\hline
    Case I & 2.08062 & 0.496059 & 1\\
    \hline
    Case II & 0.971677 & 0.500937 & 1\\
    \hline
    Case III & 0.456169 & 0.505068 & 1 \\
    \hline
\end{tabular}
\par\vspace*{6pt}\par
\begin{tabular}{|p{2cm}|p{1.5cm}|p{1.5cm}|p{1.5cm}|p{1.5cm}|}
    \hline
     New Model & $a$ & $b$ & $c$ \\
    \hline\hline
    Case I & 2.0996 & 0.469803 & 1.0151  \\
    \hline
    Case II & 0.980631 & 0.474417 & 1.01525 \\
    \hline
    Case III & 0.460408 & 0.478327 & 1.01538 \\
    \hline
\end{tabular}
\end{table}
For the sound channel, the new model provides a noticeably better fit than the old one, as can be seen in Figure \ref{fig:fitsound}. In this plot, we consider small attenuation factor $2\pi \bar{\Gamma}_s=0.002$ and vary the speed of sound. In the new model, as can be seen in Table \ref{table3}, the scaling exponent $b$ for the gravitational sound channels decreases as $v_s$ increases, indicating that the exponential contribution becomes more dominant at late times, while the power-law prefactor is progressively suppressed.\\
\indent As $v_s$ increases, the dynamics gradually shift away from the diffusive regime and exhibit stronger propagating behavior, with a progressively suppressed power-law prefactor at the late times. This result can be seen in a more detail in Figure \ref{fig:canbfit}.The parameter $c$ governs the characteristic time scale beyond which the exponential envelope becomes the dominant contribution, specifically for $t_0 \gtrsim \frac{1}{2 c \pi T}$. As $c$ increases along $v_s$, the ANEC is correspondingly more rapidly dominated by the exponential factor.\\
Furthermore, we include the $R^2$ values of the fits in Table \ref{tab:rsquare} in order to quantitatively assess the difference between the old and the new fitting models. For sound speeds close to the Schwarzschild value, $v_s \simeq v_{\text{Schw.}}$, both models provide comparably good fits. However, as the sound speed is increased, the $R^2$ values of the new model approach unity, while those of the old model gradually drift away from unity. This indicates that, although both models tend to show diffusive behavior at low sound speed, the new model accommodates the dependence on $v_s$ more reliably at larger values.

\begin{table}[ht]
\centering
\caption{\label{table3}The values of the fitting parameters $\{a,b,c\}$ for both Old and New models in gravitational sound channels, where $v_{\text{Schw.}}=\frac{1}{\sqrt{3}}$ is the Schwarzschild-AdS$_5$ sound speed. Here we fix $2\pi \bar{\Gamma}_s=0.002$ and $f_s=\frac{1}{2}$}
\vspace{2mm}
\begin{tabular}{|p{2.2cm}|p{2.2cm}|p{1.5cm}|p{1.5cm}|p{1.5cm}|}
    \hline
         Old Model & $a$ & $b$ & $c$ \\
         \hline\hline
         $v_s=v_{\text{Schw.}}$ & 0.000213639 & 0.467182 & 1\\
         \hline
         $v_s=1.2v_{\text{Schw.}}$ & 0.000434158 & 0.761162 & 1\\
         \hline
         $v_s=1.4v_{\text{Schw.}}$ & 0.0006412 & 1.01994 & 1\\
         \hline
         $v_s=1.6v_{\text{Schw.}}$ & 0.000729501 & 1.1272 & 1\\
         \hline
    \end{tabular}
    \par\vspace*{6pt}\par
    \begin{tabular}{|p{2.2cm}|p{2.2cm}|p{1.5cm}|p{1.5cm}|p{1.5cm}|}
    \hline
         New Model & $a$ & $b$ & $c$ \\
         \hline\hline
         $v_s=v_{\text{Schw.}}$ & 0.000222639 & 0.347394 & 1.06855\\
         \hline
         $v_s=1.2v_{\text{Schw.}}$ & 0.000512062 & 0.28601 & 1.27535\\
         \hline
         $v_s=1.4v_{\text{Schw.}}$ & 0.000862617 & 0.173482 & 1.49971\\
         \hline
         $v_s=1.6v_{\text{Schw.}}$ & 0.00107587 & 0.032811 & 1.66428\\
         \hline
    \end{tabular}
\end{table}
\begin{table}[]
    \caption{$R^2$ values for the fits obtained using the old and the extended (new) fitting models for different values of $v_s$.}
    \label{tab:rsquare}
    \centering
\begin{tabular}{|c|c|c|c|c|}
\hline
     $v_s$& $v_{\text{Schw.}}$ & $1.2v_{\text{Schw.}}$&$1.4v_{\text{Schw.}}$ &$1.6v_{\text{Schw.}}$ \\
     \hline\hline
     $R^2_{\text{Old}}$ & 0.992481 & 0.984236 &0.968501&0.948\\
     \hline
     $R^2_{\text{New}}$ & 0.992675&0.99133&0.992913&0.997332\\
     \hline
\end{tabular}
\end{table}
\section{Remarks on the Localized Perturbations}
In the previous section, we showed that the ANEC for both shear and sound channel perturbations exhibits a power-law scaling at late times. On the other hand, at early times or as soon as the perturbations are inserted at time $t_0$, the wormhole immediately becomes traversable as $\langle\hat{T}_{VV}^{(2)}\rangle$ is negative right after $V=V_0$ (see Figures \ref{fig:case1shear} and \ref{fig:soundmodesanecplot}). However, this might not be the case when we consider a local setting in the double trace deformation, such that the one considered in \cite{Couch_2020}. In this section, without performing an explicit analytical computation, we provide a brief discussion of how localized perturbations could alter the traversability of the wormhole.\\
\indent Instead of inserting the double trace deformation at time $t_0$ via Dirac delta function and integrating throughout all space, we could insert a local perturbation at time $t_0$ and at position $\vec{x}_0$ at both asymptotic boundaries. This could be achieved by choosing the function $A(t_1,\vec{x}_1)$ in the coupling function $Q^{\mu\nu\alpha\beta}(t_1,\vec{x}_1)$ as
\begin{align}
A(t_1,\vec{x}_1)&=\delta\bigg(\frac{2\pi}{\beta}\big((t_1-t_0)-|\vec{x}_1-\vec{x}_0|/v_B\big)\bigg)\\\nonumber
    &=\tilde{V}_0\delta(V_1-\tilde{V}_0),
\end{align}
where $\tilde{V}_0\equiv V\big|_{t=\tilde{t}_0}$ with $\tilde{t}_0\equiv t_0+|\vec{x}_1-\vec{x}_0|/v_B$. Here $v_B$ denotes the ``butterfly velocity", which indicates how fast a local information could scramble in a certain model \cite{Shenker_2014,Roberts_2015,Roberts_2016,Jahnke_2018,Prihadi_2024,Wang2025}. By choosing this function, we assume that the gravitational perturbation spread throughout the system at speed $v_B$ by placing them at the ``butterfly cone" surface.
\indent Using this choice, the first-order Green's function for gravitational sound channels is now given by
\begin{widetext}
\begin{align}
    G_{tt,tt}^{(1)}(\textbf{x};\textbf{x}^\prime)=\frac{3a}{2\pi T^3}\int d^3 x_1&\frac{d^3k}{(2\pi)^3}\frac{d^3k^\prime}{(2\pi)^3}\frac{VV^\prime\Theta(V-\tilde{V}_0)}{(1+V^\prime\tilde{V}_0)(V/\tilde{V}_0-1)}\frac{kk^\prime e^{i\vec{k}^\prime\cdot(\vec{x}^\prime-\vec{x}_1)}e^{i\vec{k}\cdot(\vec{x}-\vec{x}_1)}}{\sinh\big(\frac{k^\prime v_s}{2T}\big)}\\\nonumber
    &\times\Re(\mathcal{Q}(V^\prime,\tilde{V}_0,k))\Im(\mathcal{R}(V,\tilde{V}_0,k^\prime)).
\end{align}
We can also write the expectation value $\langle\hat{T}_{VV}^{(2)}\rangle$, although it is not integrated over all spatial $\vec{x}_1$ coordinates, as
\begin{align}\label{TVVtilde}
    8\pi G_{\text{N}}&\langle\hat{T}_{VV}^{(2)}\rangle=\frac{3a}{2^6\pi^3 T^6}\int d^3x_1\frac{d^3k}{(2\pi )^3}\frac{d^3k^\prime}{(2\pi)^3}\frac{-kk^\prime\vec{k}\cdot\vec{k}^\prime}{\sinh\big(\frac{k^\prime v_s}{2T}\big)}\frac{\Re(\mathcal{Q}(V^\prime,\tilde{V}_0,k))\Im(\mathcal{R}(V,\tilde{V}_0,k^\prime))}{P(V,\tilde{V}_0)}. 
\end{align}
\end{widetext}
\indent We would like to give comments to the expectation value of the energy-momentum tensor at both late and early time. At a late time when $V\gg1$, the dependence on $\tilde{V}_0$ in Eq. \eqref{TVVtilde} cancels out, and now the result does not care about the initial location of the perturbation anymore. This is because, as $(t-t_0)\gg|\vec{x}_1-\vec{x}_0|/v_B$, the information has already spread to some large enough region, and now we are inside the "butterfly cone" \cite{jahnke2019recent}. However, at an early time, the situation is different.\\
\indent At early time, right after the insertion time $t\sim t_0$, $\langle\hat{T}_{VV}^{(2)}\rangle$ diverges and negative as $V/V_0\rightarrow1$ while before that, the energy-momentum tensor is positive-valued. However, for localized perturbations, $V/\tilde{V}_0\rightarrow1$ is only achieved after $t-t_0\sim |\vec{x}_1-\vec{x}_0|/v_B$. This implies that one must wait for a while until the perturbation has sufficiently scrambled over the relevant region, or until the system enters the butterfly cone. Before entering the butterfly cone, the energy-momentum tensor is given by
\begin{align}
    8&\pi G_{\text{N}}\langle\hat{T}_{VV}^{(2)}\rangle\approx\frac{3av_s}{2^7\pi^4T^6}\int d^3x_1\frac{d^3k}{(2\pi)^3}\frac{d^3k^\prime}{(2\pi)^3}\\\nonumber
    &\times\frac{e^{i(\vec{x}-\vec{x}_1)\cdot(\vec{k}+\vec{k}^\prime)}}{\sinh\big(\frac{k^\prime v_s}{2T}\big)}k^2k^\prime(-\vec{k}\cdot\vec{k}^\prime)Ve^{-2\pi T(t_0+|\vec{x}_1-\vec{x}_0|/v_B)},
\end{align}
as we may approximate
\begin{align}
    P(V,\tilde{V}_0)&\approx-1,\\
    \Re(\mathcal{Q}(V,\tilde{V}_0,k))&\approx1,\\
    \Im(\mathcal{R}(V,\tilde{V}_0,k))&\approx-\frac{kv_s}{2\pi T}\frac{V}{\tilde{V}_0}.
\end{align}
One can see that it is now suppressed by $e^{-2\pi T|\vec{x}_1-\vec{x}_0|/v_B}$, and hence small, up until the time reach $t-t_0\sim |\vec{x}_1-\vec{x}_0|/v_B$. Therefore, although $t>t_0$, the energy-momentum tensor should still be positive when the system is still outside the butterfly cone ($V/\tilde{V}_0<1$).
\section{Summary and Discussions}\label{Section6}
This work extends previous studies \cite{GaoJW, Ahn} on traversable wormholes via double trace deformation. Instead of using scalar or vector field, we perturb the AdS$_5$ black brane background with gravitational/tensor perturbations that are dual with energy momentum tensor operators on its boundary. The approach is carried out through linear perturbation analysis in the shear and sound channels.\\\\
\textbf{Gravitational shear channel.} We consider three different coupling functions, $Q^{\mu\nu\alpha\beta}(t,\vec{x})$, for the shear channel, denoted as case I, case II, and case III. We find that each case results in a different opening of the wormhole, $\Delta U^{(2)}$, as seen in Figure \ref{fig:case1shearANECintegral}, with the order of traversability from the largest to the smallest being $tx-tx, tx-zx,$ and $zx-zx$. We know that $tx-tx$ is a momentum density-momentum density correlation. This type of coupling gives the strongest traversability compared with momentum density-momentum flux and momentum flux-momentum flux interactions.\\
\indent We also note that, in subsection \ref{greenfunctionshearchannelcalc}, we generalize the tensorial diffusion constant, $\mathcal{D}_T$, for an arbitrary function $f(u)$ that can recover $\mathcal{D}_T=1/4\pi T$ for Shwarzschild AdS$_5$ blackhole which has $f(u)=1-u^2/u_H^2$, which is consistent with the results of previous studies \cite{natsuume2016adscftdualityuserguide, Policastro:2002se}. Thus, one can verify if we could get different diffusion constants for different forms of $f(u)$ that could possibly affect the the wormhole opening. If one apply $f(u)$ for a black hole with momentum relaxation \cite{Ahn}, one could obtain $\mathcal{D}_T=\frac{1}{2}\mathcal{D}_c$.\\\\
\textbf{Gravitational sound channel.} For the sound channel, we find that its traversability is smaller than that of all the shear channels. Nevertheless, we observe some interesting findings regarding this channel. When plotting the ANEC integral, as seen in Figure \ref{fig:case1shearANECintegral} versus Figure \ref{fig:soundmodesanecplot}, the turning point of the graph, corresponds to maximal traversability, is different. For the shear channel case, the turning point is shown when $V_0=1$, equivalent with $t_0=0$. However, with the sound channel, the wormhole becomes most traversable when the insertion time is $t_0<0$, indicating that we need to activate the perturbations earlier.\\
\indent This raises further questions regarding the underlying mechanism. We speculate that this comes from the propagating nature of the gravitational sound channel. At low speed, the perturbations behaves more like a diffusive probes similar with $U(1)$ conserved current operators where it is found that maximal traversability in obtained when $t_0=0$. As the sound speed becomes larger, however, the response becomes less diffusive and prefers propagating wave-like properties. In that case, as can be seen in Figure \ref{fig:variasiumax}, the time for maximal traversability is shifted to $t_0<0$. This behavior remains speculative at this stage. Nevertheless, a more detailed investigation involving localized perturbations may lead to conclusions aligned with the results of \cite{Cheng_2021}, where distinct scrambling features were observed depending on the ratio between $v_s$ and the butterfly velocity $v_B$. Clarifying this connection gives an interesting open problem that we leave for future work.\\
\indent Another aspect of the sound channel perturbations is the influence of the attenuation constant on the wormhole traversability. The result can be seen in Figure \ref{fig:varyattenuation}, where a higher attenuation constant leads to less traversability for early $t_0$ while it increases traversability at later times. This seems to contradict the behavior of the ANEC in the diffusive modes. In the sound channel, however, the relevant quantity controlling traversability is the ratio $\bar{\Gamma}_s/v_s^2$. Therefore, the effect between $\bar{\Gamma}_s$ and $v_s$ compete with each other at both early and late insertion times, leading to the result found in Figure \ref{fig:varyattenuation}. Furthermore, once the perturbations become superluminal, as can be seen in Figure \ref{fig:superluminal}, the opening of the wormhole becomes too transient such that the wormhole becomes non-traversable at late insertion time, although the turning point returns to $t_0\approx0$.\\\\
\textbf{Power-law remnants of the ANEC at late times}. In Section \ref{section4}, we perform data fitting and find that the ANEC for both the shear and sound channels can be well described by
\begin{equation}
\mathcal{A}_{\text{new}}(V_0)=-\frac{V_0}{V_0^{(c+1)}+1}\frac{a}{[\ln(V_0+1/V_0)]^b},
\end{equation}
where $\{a,b,c\}$ are fitting parameters. At sufficiently large time, the behavior of the ANEC is therefore governed by the numerical values of the parameters $b$ and $c$, which control whether the decay is purely exponential behavior or exponential suppression modified by an additional power-law factor. At late times $(t_0\gg1)$, the expression reduces to Eq. \eqref{powerlawtail}.\\
\indent In the shear channel, we obtain results comparable to those reported in \cite{Ahn}, with the parameter $c$ remaining close to unity. Consequently, the fitting procedure effectively reduces to the previously employed model (see Table \ref{table2}). In the sound channel, although the same functional form continues to provide an adequate description of the data, the fitted parameters display a systematic numerical shift: $c$ deviates from unity and $b$ decreases as the speed of sound is increased, as summarized in Table \ref{table3} and illustrated in Figure \ref{fig:canbfit}. This behavior implies that the relative contribution of the power-law factor diminishes at larger values of the sound speed, and the late-time behavior increasingly dominated by purely exponential decay.\\
\indent This trend suggests that sound channel interpolates between diffusive and propagating behavior as the relative values of the sound speed and the attenuation constant are varied. In this sense, although the functional form governing the late-time behavior of the sound and shear channels is qualitatively similar, our results indicate a quantitative shift from diffusion-dominated behavior toward propagation-dominated behavior, reflected in the changing importance of the power-law factor as controlled by the fitted parameters $b$ and $c$.\\\\
\textbf{Bound on information transfer}. An important distinction between the gravitational vector type perturbation, $\mathfrak{h}_{tx}$ and $\mathfrak{h}_{zx}$, versus pure vector perturbation $A_{\mu}$ is that the energy-momentum tensor $\langle T_{VV}\rangle$ for vector fields depends solely on the gauge fields $A_V$, while the gravitational contribution explicitly involves the gravitational constant $G_{\text{N}}$. The appearance of $G_{\text{N}}$ encodes the backreaction of spacetime geometry, shifting the horizon and allowing the signal/information to be sent. It also underlines gravitational nature of the interaction, distinguishing it from other gauge interactions.\\
\indent One can calculate the upper bound of how much information can be transferred through the wormhole  opening caused by gravitational interaction, denoted as $N_{\text{bits}}$ using the uncertainty principle \cite{Ahn,Freivogel_2020},
\begin{equation}
    p_U^\text{each}\Delta U_\text{each}\gtrsim  1.
\end{equation}
The number of bits $N_{\text{bits}}$ is upper-bounded by $\Delta U$ times momentum of the shockwaves $p^U$, which is roughly equal or less than $(r_H^{d-1})/G_{\text{N}}$ in a so called probe limit. This limit is obtained from the assumption that the backreaction from a particle with positive energy that is sent through the wormhole is substantially smaller than the backreaction from gravitational perturbations that open the wormhole. In our case, $\Delta U$ is proportional to $8\pi G_{\text{N}}\int\langle\hat{T}_{VV}^{(2)}\rangle dV$, while $8\pi G_{\text{N}}\langle\hat{T}_{VV}^{(2)}\rangle\sim\mathcal{O}(h_{MN}^2)$ (see Eq. \eqref{EOMgamma}).  Therefore, to have $N_{\text{bits}}\sim\mathcal{O}(1)$, the gravitational perturbations should be at least $h_{MN}\sim\mathcal{O}(\sqrt{G_{\text{N}}})$, whereas $h_{MN}\sim\mathcal{O}(G_{\text{N}})$ leads to $N_{\text{bits}}\sim\mathcal{O}(G_{\text{N}})$, which is vanishing in a semiclassical limit. The behavior of $h_{MN}\sim\mathcal{O}(\sqrt{G_{\text{N}}})$ is expected in the perturbative quantum gravity (see, for example \cite{Bern_2002,lvarez_2020}).\\
\indent In the end of Section \ref{section2} we  discussed although the deformation is irrelevant, it nevertheless can  render wormhole traversable under hydrodynamic regime. The physical constraint is the ability to send $N_{\text{bits}}$ of information through the wormhole within the probe limit, where the shock waves momentum satisfies  $p^U \lesssim r_H^{d-1}/G_{\text{N}}$. Physically, this means that the signal backreaction is substantially smaller than the double trace deformation. Our calculations using the low-energy hydrodynamic limit confirm that this condition is consistently satisfied for $h_{MN} \sim O(\sqrt{G_{\text{N}}})$.
\\\\
\textbf{Future works}. So far, the construction of traversable wormholes with double trace deformation has been accomplished by using bosonic operators such as scalar, vector, and, in this paper, tensor operators. Then it is interesting to see if this procedure can be developed for fermionic operators. Unlike scalar, vector, and tensor, there are some subtleties when choosing the boundary condition in fermionic interaction. This is because when the bulk action vanishes at the boundary, we can add boundary terms to the action that satisfy the variational principles, and different boundary terms can lead to different conformal field theories \cite{Laia2011AHF}. Therefore, one could produce a different kind of deformation. In other contexts, the lack of fermionic superradiance is an advantage to construct traversable wormholes in a Kerr black hole \cite{Iyer,Martellini}. The bosonic superradiance prevents wormholes from being defined at the off-axis region \cite{Bilotta}, while fermions can avoid this problem. By having traversable wormholes from more physical black holes (especially near the extreme limit), it can open up the possibility of observing wormholes in the future. Wormhole traversability from fermionic operators will be presented in the near future. \\
\indent Another interesting aspect to explore is the study of traversable wormhole for various black hole backgrounds, especially involving scalar or vector hair, which have attracted significant interest until very recently \cite{Hartnoll_2021,Cai_2022,Cai2023towards,Prihadi2025a,Prihadi2025b,xu2025black}. It is shown that black hole parameters can be influenced by the value of the scalar field at the asymptotic AdS boundary, which corresponds to a deformation in the boundary CFT. One might also expect that the boundary deformation could influence the diffusion constant or the speed of sound of the theory. Furthermore, the use of rotating shock waves \cite{Malvimat2022,Prihadi_2023,Prihadi_2024,Prihadi2025b} might also influence the traversability of a wormhole, especially in rotating backgrounds \cite{Bilotta}, or even from its quantum counterpart called the Exotic Compact Object (ECO) \cite{Djogama_2024}. The angular momentum of the shock waves gives an enhanced upper bound for the Lyapunov exponent in the context of gravitational scrambling dynamics.\\
\indent Lastly, this work may also be connected to inflationary cosmology, where it suggests that primordial gravitational perturbations, generated during inflation epochs, originated from quantum fluctuations in the early universe \cite{SatoYokoyama2015, Chang_2023}. Analogous to the tensor perturbation, cosmological perturbations can also be classified as scalar, vector, or tensor modes depending on their behavior under spatial coordinate transformations. \cite{Chang_2023}. Furthermore, in the aforementioned work, gravitational waves are represented as second-order perturbations. In our study, the metric perturbations are also treated at second order, which motivates us to find out if such gravitational wave perturbations may create a traversable wormhole. Additionally, a recent proposal suggests the connection between inflationary cosmology and Anti-de Sitter wormholes, initiated by an effort to address the problem of the initial conditions for inflation \cite{PanosBetziosPhysRevLett.133.021501}.
\section*{Acknowledgements}
FK would like to thank the Ministry of Primary and Secondary Education (Kemendikdasmen) for the financial support through the Beasiswa Unggulan program and Badan Riset dan Inovasi Nasional (BRIN) for partial financial support through the Research Assistant program. HLP would like to thank Badan Riset dan Inovasi Nasional (BRIN) for financial support through the Post-doctoral program. DD is supported by the APCTP (YST program) through the Science and Technology Promotion Fund and Lottery Fund of the Korean Government and the Korean Local governments, Gyeongsangbuk-do Province, and Pohang city. FPZ would like to thank riset PPMI, Fakultas Matematika dan Ilmu Pengetahuan Alam, Institut Teknologi Bandung and the Ministry of Higher Education, Science, and Technology (Kemendiktisaintek) for financial support.
\section*{Appendix A: The Energy-Momentum Tensor of the Gravitational Fields}\label{appendixA}
In this appendix, we show how we obtain the expression of $T_{VV}^{(2)}$ for both shear and sound channels as in Eqs. \eqref{Tvv} and \eqref{Tvvsound}, respectively. For the shear channels, we calculate the $VV$ component of the Einstein's equation
\begin{equation}
    R_{MN}[g]-\frac{1}{2}R[g]g_{MN}-6g_{MN}=0,
\end{equation}
for the metric components $g_{MN}$ given by Eq. \eqref{perturbedvector}. We rescale $h_{MN}\rightarrow\alpha h_{MN}$ and consider the factor $\alpha$ as the perturbation parameter, and we expand the equations of motion up to order $\mathcal{O}(\alpha^2)$. The energy-momentum tensor is given by the quadratic order of the Einstein's equation, as shown in Eq. \eqref{Tvvkuadrat}.We then take the low-temperature limit $\beta\rightarrow \infty$, ignoring the subleading terms, and take the near-horizon limit $U\rightarrow0$.\\
\indent The $\mathcal{O}(\alpha^2)$ terms of the Einstein's equation near $U\rightarrow0$ for all orders of $\beta$ can be written as
\begin{align}
    &\frac{\beta^2}{32\pi^2V^2}(\partial_z\mathfrak{h}_{tx})^2-\frac{\beta}{4\pi V}(\partial_z\mathfrak{h}_{xz}\partial_V\mathfrak{h}_{tx}\\\nonumber&+\mathfrak{h}_{xz}\partial_z\partial_V\mathfrak{h}_{tx})+\frac{g_{xx}}{2Vg_{tt}}(\mathfrak{h}_{tx}\partial_V\mathfrak{h}_{tx}+V\mathfrak{h}_{tx}\partial_V^2\mathfrak{h}_{tx})\\\nonumber  &+\mathfrak{h}_{xz}\partial_V^2\mathfrak{h}_{xz}+\frac{1}{V}\mathfrak{h}_{xz}\partial_V\mathfrak{h}_{xz}+\frac{1}{2}(\partial_V\mathfrak{h}_{xz})^2.
\end{align}
In the hydrodynamic limit, we consider the case with low temperature and near the horizon. The terms proportional to $\frac{1}{g_{tt}}$ may diverge near the horizon. However, since we take the $\beta \to \infty$ limit first, those terms become subleading and can thus be neglected in this regime. Consequently, the energy-momentum tensor remains finite in the hydrodynamic limit, as intended.\\
\indent For the sound channels, we do similar calculations for the metric given by Eq. \eqref{perturbedmetricsound}. Near the horizon, explicit calculations of the $\mathcal{O}(\alpha^2)$ terms of the Einstein's equation give us
\begin{align}
    &\frac{\beta^2}{64\pi^2V^2}\bigg[\frac{g_{xx}}{g_{tt}}(\partial_z\mathfrak{h}_{tt})^2+2\frac{g_{xx}}{g_{tt}}\mathfrak{h}_{tt}\partial_z^2\mathfrak{h}_{tt}\\\nonumber
    &\;\;\;\;\;\;\;\;\;\;\;\;\;\;\;\;\;\;-\partial_z\mathfrak{h}_{tt}\partial_z\mathfrak{h}_{zz}+2\partial_z\mathfrak{h}_{tt}\partial_z\mathfrak{h}_{xx}\\\nonumber&\;\;\;\;\;\;\;\;\;\;\;\;\;\;\;\;\;\;-2\mathfrak{h}_{tt}\partial_z^2\mathfrak{h}_{zz}-4\mathfrak{h}_{tt}\partial_z^2\mathfrak{h}_{xx}\bigg]\\\nonumber
    &+\frac{\beta}{16\pi V}\bigg[\frac{3g_{xx}}{g_{tt}}\partial_z\mathfrak{h}_{zt}\partial_V\mathfrak{h}_{tt}-\frac{g_{xx}}{g_{tt}}\partial_z\mathfrak{h}_{tt}\partial_V\mathfrak{h}_{zt}\\\nonumber&\;\;\;\;\;\;\;\;\;\;\;\;\;\;\;\;\;\;+4\partial_z\mathfrak{h}_{xx}\partial_V\mathfrak{h}_{zt}-2\partial_z\mathfrak{h}_{zt}\partial_V\mathfrak{h}_{zt}\\\nonumber&
    \;\;\;\;\;\;\;\;\;\;\;\;\;\;\;\;\;\;+2\frac{g_{xx}}{g_{tt}}\mathfrak{h}_{zt}\partial_V\partial_z\mathfrak{h}_{tt}+2\frac{g_{xx}}{g_{tt}}\mathfrak{h}_{tt}\partial_V\partial_z\mathfrak{h}_{zt}\\\nonumber&\;\;\;\;\;\;\;\;\;\;\;\;\;\;\;\;\;\;-4\mathfrak{h}_{zz}\partial_V\partial_z\mathfrak{h}_{zt} \bigg]\\\nonumber
    &+\frac{1}{V}\mathfrak{h}_{xx}\partial_V\mathfrak{h}_{xx}-\frac{g_{xx}}{2g_{tt}}\partial_V\mathfrak{h}_{tt}\partial_V\mathfrak{h}_{xx}+\frac{1}{2}(\partial_V\mathfrak{h}_{xx})^2\\\nonumber
    &+\frac{g_{xx}}{2Vg_{tt}}\mathfrak{h}_{zt}\partial_V\mathfrak{h}_{zt}+\frac{1}{2V}\mathfrak{h}_{zz}\partial_V\mathfrak{h}_{zz}-\frac{g_{xx}}{4g_{tt}}\partial_V\mathfrak{h}_{tt}\partial_V\mathfrak{h}_{zz}\\\nonumber
    &+\frac{1}{4}(\partial_V\mathfrak{h}_{zz})^2+\mathfrak{h}_{xx}\partial_V^2\mathfrak{h}_{xx}\\\nonumber
    &+\frac{g_{xx}}{2g_{tt}}\mathfrak{h}_{zt}\partial_V^2\mathfrak{h}_{zt}+\frac{1}{2}\mathfrak{h}_{zz}\partial_V^2\mathfrak{h}_{zz}.
\end{align}
Here, even $\beta^2$ terms also contains $\frac{1}{g_{tt}}$ and thus diverge at the horizon. In order to overcome this, we note that there is a discrepancy between the definition of $\mathfrak{h}_{tt}$ and $H_t=g^{tt}h_{tt}\neq\mathfrak{h}_{tt}$ that is used in \cite{MATSUO2009593}. This is in contradiction with other channels where $\mathfrak{h}_{\mu\nu}=g^{xx}h_{\mu\nu}$. We choose to perform a rescaling so that $\mathfrak{h}_{tt}\rightarrow\sqrt{\frac{g_{tt}}{g_{xx}}}\mathfrak{h}_{tt}$. This rescaling effectively absorbs the redshift factor associated with the vanishing of $g_{tt}$ near the horizon. Physically, it corresponds to expressing the time-time component of the perturbation in terms of quantities measured by a locally inertial observer. After the rescaling and ignoring the subleading terms in $\beta\rightarrow\infty$, only the terms in Eq. \eqref{Tvvsound} appear in the quadratic energy-momentum tensor of the gravitational sound channels.
\section*{Appendix B: Relation between Retarded Green's Function and Wightman Function}\label{appendixB}
Retarded Green's Function between operator A and operator B denotes as \cite{ACavalloPhysRevE.77.051110}
\begin{align}
    G_{AB}^\text{ret.}(t,t^\prime)&=-i\langle[\hat{A}(t),\hat{B}(t^\prime)]\rangle\\\nonumber
    &=-\frac{i}{Z(\beta)}(\text{Tr}\{e^{-\beta\hat{H}}\hat{A}(t)\hat{B}(t^\prime)\}\\\nonumber
    &\;\;\;\;\;\;\;\;\;\;\;\;\;\;\;\;\;\;-\text{Tr}\{e^{-\beta\hat{H}}\hat{B}(t^\prime)\hat{A}(t)\}).
\end{align}
In this case, we use $Z\equiv \text{Tr}\{ e^{-\beta H}\}$, $\langle \hat{X}\rangle=\text{Tr}(e^{-\beta H} \hat{X})/\text{Tr} \{e^{-\beta H}\}$. By expanding the quantum state using spectral decomposition with $|n\rangle$ is the energy eigenstate and using Heisenberg representation of the operator, we get
\begin{align}
    G^{\text{ret.}}_{AB}&=-\frac{i}{Z}\bigg(\sum_{n,m}A_{nm}B_{mn} e^{-\beta E_n}e^{i(E_n-E_m)(t-t^\prime)}\\ & \nonumber\;\;\;\;\;\;\;\;-\sum_{n,m}B_{nm}A_{mn} e^{-\beta E_n}e^{-i(E_n-E_m)(t-t^\prime)} \bigg)\\
    &= -\frac{i}{Z}\sum_{n,m}A_{nm} B_{mn} e^{-\beta E_n}e^{i(E_n-E_m)(t-t^\prime)}\\\nonumber
&\;\;\;\;\;\;\;\;\;\;\;\;\;\;\;\;\;\;\;\;\times (1-e^{\beta(E_n-E_m)})
\end{align}
where $A_{nm}=\langle n|A|m\rangle$, and we also exchange the dummy indices $m$ and $n$ to get from the first line to the second line. Fourier transforming the Green's function we get
\begin{align}
    G_{AB}^\text{ret.}(\omega)=&\int d(t-t^\prime)G_{AB}^\text{ret.}(t,t^\prime)e^{i\omega(t-t^\prime)}\\\nonumber
    =&-\frac{2\pi i}{Z}(1-e^{-\beta\hat{H}})\sum_{nm}A_{nm}B_{mn}e^{-\beta\hat{H}}\\\nonumber
    &\;\;\;\times\delta(\omega-(E_m-E_m)).
\end{align}
On the other hand, Wightman function is given by \cite{Hsiang_2024, ACavalloPhysRevE.77.051110}
\begin{align}
    G^W_{AB}(t,t^\prime)&=-i\langle A(t)B(t^\prime)\rangle \\
    =-&\frac{i}{Z}\sum_{n,m}A_{nm}B_{mn}e^{-\beta E_n}e^{i(E_n-E_m)(t-t^\prime)},
\end{align}
where we have used the same treatments as we did to $G^{ret}_{AB}$. Similarly, Fourier transforming we get Wightman function in Fourier space, we get
\begin{align}
    G^W_{AB}(\omega)=&-\frac{i}{Z}\sum_{n,m}A_{nm}B_{mn}e^{-\beta E_n}\\&\nonumber\times2\pi \delta(\omega-(E_m-E_n)).
\end{align}
From here, we get the relation between retarded Green's function and Wightman function, that is given by
\begin{equation}
    G^W_{AB}(\omega)=i\frac{e^{\beta\omega}}{e^{\beta\omega}-1}\text{Im}G_{AB}^{\text{ret.}}(\omega).
\end{equation}
\section*{Appendix C: Solutions to the Equations of Motion in Shear Channels}\label{appendixC}
The equation of motion for $h_t(r)$ in the static and spherically-symmetric AdS black hole background is given by 
\begin{align}
    &h_t^{\prime\prime\prime}(u)-h_t^{\prime\prime}(u)\partial_u\log(u^2g_{uu})\\\nonumber
    &-ug_{uu}\bigg(\frac{\omega^2g^{tt}}{u}+k^2\bigg)h_t^\prime(u)+\frac{1}{u}\partial_u\log(u^2g_{uu})h_t^\prime(u)=0.
\end{align}
Using infalling boundary condition at the horizon, we can write $h_t^\prime(u)=(u_H-u)^{-i\omega/\nu}F(u)$, with $\nu\equiv 2\sqrt{u_H}|F^\prime(u_H)|$. Then, the equation becomes
\begin{align}
    F^{\prime\prime}+&F^\prime\bigg\{\frac{2i\omega}{\nu(u_H-u)}-\partial_u\log(u^2g_{uu})\bigg\}\\\nonumber
    &-F\bigg\{\omega^2\bigg(g_{uu}g^{tt}+\frac{1}{\nu^2(u_H-u)^2}\bigg)+ug_{uu}k^2\bigg\}\\\nonumber
    +F&\bigg\{\frac{i\omega}{\nu}\bigg(\frac{1}{(u_H-u)^2}-\frac{\partial_u\log(u^2g_{uu})}{(u_H-u)}\bigg)\\\nonumber
    &+\frac{1}{u}\partial_u\log(u^2g_{uu})\bigg\}=0.
\end{align}
We solve this equation perturbatively by expanding $F(u)$ as
\begin{equation}
    F(u)=F_0(u)+\omega F_1(u)+k^2G_1(u)+\mathcal{O}(\omega^2,\omega k^2,k^4).
\end{equation}
\indent At order $\mathcal{O}(1)$, the equation is given by
\begin{equation}
    F_0^{\prime\prime}-\bigg(F_0^\prime-\frac{F_0}{u}\bigg)\partial_u\log(u^2g_{uu})=0,
\end{equation}
which has the solution in the form of
\begin{equation}
    F_0=uC_1\int^ug_{uu}(u_1)du_1+uC_0.
\end{equation}
Near the horizon, $F_0(u)$ behaves as
\begin{align}
    F_0&\approx u_HC_1\int^u\frac{du_1}{4u_H^2|F^\prime(u_H)|(u_H-u_1)}+u_HC_0\\\nonumber
    &=\frac{-C_1}{4u_H|F^\prime(u_H)|}\log|u-u_H|+u_HC_0.
\end{align}
Regularity at the horizon gives us $C_1=0$ and
\begin{equation}
    F_0(u)=uC_0.
\end{equation}
\indent At order $\mathcal{O}(\omega)$, the equation is given by
\begin{align}
    F_1^{\prime\prime}+\bigg(-F_1^\prime&+\frac{F_1}{u}\bigg)\partial_u\log(u^2g_{uu})\\&=-\frac{iuC_0}{\nu}g_{uu}\partial_u\bigg(\frac{1}{g_{uu}(u_H-u)}\bigg).
\end{align}
which has the solution in the form of
\begin{align}
    F_1=&u\int^u\bigg(\frac{-iC_0/\nu}{u_H-u_1}+C_1g_{uu}(u_1)\bigg)du_1+uC_2.
\end{align}
Near the horizon, this function becomes
\begin{align}
    F_1(u)\approx& u\int^u\bigg(\frac{-iC_0/\nu}{u_H-u_1}\\&\nonumber+\frac{C_1}{4u_H^2|F^\prime(u_H)|(u_H-u_1)}\bigg)du_1+uC_2\\\nonumber
    =&\nonumber r_H\bigg(\frac{iC_0}{\nu}-\frac{C_1}{4u_H^2|F^\prime(u_H)|}\bigg)\log|u_H-u|\\&\nonumber+u_HC_2.
\end{align}
A solution that is regular at the horizon have
\begin{equation}
    C_1=\frac{i4u_H^2|F^\prime(r_H)|C_0}{\nu}.
\end{equation}
Therefore, by absorbing the integration constant $C_2$ into the integration interval, we get
\begin{align}
    F_1(u)=&\frac{iC_0}{\nu}u\int_{u_H}^u\bigg(4u_H^2|F^\prime(u_H)|g_{uu}(u_1)-\frac{1}{u_H-u_1}\bigg)du_1\\\nonumber
    \equiv&-iC_0H_1(u).
\end{align}
The function
\begin{equation}
    H_1(u)=\frac{u}{\nu}\int_{u}^{u_H}\bigg(4u_H^2|F^\prime(u_H)|g_{uu}(u_1)-\frac{1}{u_H-u_1}\bigg)du_1,
\end{equation}
is finite and positive at the boundary $u=0$.\\
\indent At order $\mathcal{O}(k^2)$, the equation is given by
\begin{equation}
    G_1^{\prime\prime}+\bigg(-G_1^\prime+\frac{G_1}{u}\bigg)\partial_u\log(u^2g_{uu})-u^2g_{uu}C_0=0,
\end{equation}
which has the solution in the form of
\begin{equation}
    G_1=u\int^u g_{uu}(u_1)\bigg(\frac{C_0u_1^2}{2}+C_1\bigg)du_1+uC_2.
\end{equation}
Near the horizon, this solution behaves as
\begin{align}
    G_1(u)\approx& u_H\int^u\frac{-1}{4u_H^2|F^\prime(u_H)|(u-u_H)}\\\nonumber
    &\bigg(\frac{C_0u_H^2}{2}+C_1\bigg)du_1+u_HC_2\\\nonumber
    =&\frac{-1}{4u_H|F^\prime(u_H)|}\bigg(\frac{C_0u_H^2}{2}+C_1\bigg)\log|u_H-u|\\\nonumber
    &+u_HC_2.
\end{align}
Regularity at the horizon gives us
\begin{equation}
    C_1=-\frac{C_0u_H^2}{2}.
\end{equation}
Therefore, by absorbing the integration constant $C_2$ again to the integral we get
\begin{align}
    G_1(u)&=\frac{uC_0}{2}\int_{u_H}^u g_{uu}(u_1)(u_1^2-u_H^2)du_1\\&\nonumber\equiv C_0 H_2(u).
\end{align}
Now, the function
\begin{equation}
    H_2(u)=\frac{u}{2}\int_{u_H}^ug_{uu}(u_1)(u_1^2-u_H^2)du_1,
\end{equation}
is also finite and positive at the boundary $u=0$.\\
\indent By combining all of the solutions, we get
\begin{align}
    h_t^\prime(u)=C_0&(u_H-u)^{-i\omega/\nu}\big(u-i\omega H_1(u)\\\nonumber&+k^2H_2(u)
    +\mathcal{O}(\omega^2,\omega k^2,k^4)\big).
\end{align}
The constant $C_0$ can be obtained from the solution at the boundary $u=0$. The value of $h_t$ and $h_z$ at the boundary are denoted as $h_t^0$ and $h_z^0$ respectively. By taking the limit $r\rightarrow0$ in Eq. \eqref{eomht} and using the AdS boundry condition $f(0)=1$, we get
\begin{equation}
    C_0=\frac{k^2h_t^0+k\omega h_z^0}{H_1(0)(i\omega-\mathcal{D}_Tk^2)}u_H^{i\omega/\nu}+\mathcal{O}(\omega^2,\omega k^2,k^4).
\end{equation}
This is equivalent to the result obtained in \cite{Policastro:2002se} when $u_H=1$, thus in this work, we generalize the value of $u_H$.

\bibliography{bibliography.bib}
\end{document}